\documentclass[prd,aps,reprint,noeprint,superscriptaddress,amsmath, amssymb]{revtex4-1}

\usepackage{graphicx}\usepackage{dcolumn}\usepackage{bm}\usepackage{lineno}
\usepackage{epsfig}
\usepackage{multirow}
\usepackage{overpic}
\usepackage{color}
\usepackage{xspace} \usepackage{booktabs,tabularx}
\usepackage{array}
\usepackage{textcomp}
\usepackage[colorlinks,linkcolor=blue,urlcolor=blue,citecolor=blue]{hyperref}

\usepackage{cleveref}

\usepackage{dcolumn}

\newcommand{\psip}{\psi(3686)}

\newcommand{\jpsi}{J/\psi}

\newcommand{\chicz}{\chi_{c0}}

\newcommand{\llep}{\ell^+\ell^-}
\newcommand{\ee}{e^+e^-}

\newcommand{\pip}{\pi^+}
\newcommand{\pim}{\pi^-}
\newcommand{\piz}{\pi^0}
\newcommand{\pipi}{\pi^+\pi^-}
\newcommand{\pizpiz}{\pi^0\pi^0}

\newcommand{\ks}{K^0_S}
\newcommand{\kp}{K^{+}}
\newcommand{\km}{K^{-}}
\newcommand{\hc}{h_{c}}
\newcommand{\gam}{\gamma}
\newcommand{\gev}{{\rm GeV}}
\newcommand{\gevc}{{\rm GeV}/c}

\newcommand{\mev}{\rm MeV}

\newcommand{\mevcsq}{{\rm MeV}/c^{2}}

\newcommand{\lumtot}{{22.1~\rm fb^{-1}}}

\newcommand{\YI}{Y(4230)}
\newcommand{\YII}{Y(4390)}
\newcommand{\YIII}{Y(4660)}

\newcommand{\massTwo}{{\rm 4383.0\pm8.6\pm1.9}}
\newcommand{\massThree}{{\rm 4684.0\pm17.3\pm1.9}}

\newcommand{\widthTwo}{{117.4\pm20.7\pm4.8}}
\newcommand{\widthThree}{{119.5\pm47.1\pm9.1}}

 \newcommand{\BESIIIorcid}[1]{\href{https://orcid.org/#1}{\hspace*{0.1em}\raisebox{-0.45ex}{\includegraphics[width=1em]{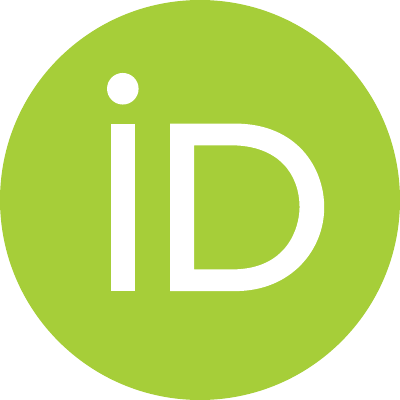}}}}

\begin{document}
\normalsize
\parskip=5pt plus 1pt minus 1pt


\title
{
  \boldmath Cross section measurement of $\ee\to\piz\piz\psip$ from $\sqrt{s}=$ 4.008~GeV to 4.951~GeV
}

\author
{
  \begin{center}
    \begin{small}
M.~Ablikim$^{1}$\BESIIIorcid{0000-0002-3935-619X},
    M.~N.~Achasov$^{4,c}$\BESIIIorcid{0000-0002-9400-8622},
    P.~Adlarson$^{81}$\BESIIIorcid{0000-0001-6280-3851},
    X.~C.~Ai$^{86}$\BESIIIorcid{0000-0003-3856-2415},
    R.~Aliberti$^{39}$\BESIIIorcid{0000-0003-3500-4012},
    A.~Amoroso$^{80A,80C}$\BESIIIorcid{0000-0002-3095-8610},
    Q.~An$^{77,64,\dagger}$,
    Y.~Bai$^{62}$\BESIIIorcid{0000-0001-6593-5665},
    O.~Bakina$^{40}$\BESIIIorcid{0009-0005-0719-7461},
    Y.~Ban$^{50,h}$\BESIIIorcid{0000-0002-1912-0374}
% ...
    \\
    \vspace{0.2cm}
    (BESIII Collaboration)\\
    \vspace{0.2cm}
    {\it
      $^{1}$ Institute of High Energy Physics, Beijing 100049, People's Republic of China\\
% ...
$^{4}$ Budker Institute of Nuclear Physics SB RAS (BINP), Novosibirsk 630090, Russia\\
% ...
$^{39}$ Johannes Gutenberg University of Mainz, Johann-Joachim-Becher-Weg 45, D-55099 Mainz, Germany\\
      $^{40}$ Joint Institute for Nuclear Research, 141980 Dubna, Moscow region, Russia\\
% ...
$^{50}$ Peking University, Beijing 100871, People's Republic of China\\
% ...
$^{62}$ Southeast University, Nanjing 211100, People's Republic of China\\
% ...
$^{64}$ State Key Laboratory of Particle Detection and Electronics, Beijing 100049, Hefei 230026, People's Republic of China\\
% ...
$^{77}$ University of Science and Technology of China, Hefei 230026, People's Republic of China\\
% ...
$^{80}$ University of Turin and INFN, (A)University of Turin, I-10125, Turin, Italy; (B)University of Eastern Piedmont, I-15121, Alessandria, Italy; (C)INFN, I-10125, Turin, Italy\\
      $^{81}$ Uppsala University, Box 516, SE-75120 Uppsala, Sweden\\
% ...
$^{86}$ Zhengzhou University, Zhengzhou 450001, People's Republic of China\\
      \vspace{0.2cm}
      $^{\dagger}$ Deceased\\
% ...
$^{c}$ Also at the Novosibirsk State University, Novosibirsk, 630090, Russia\\
% ...
$^{h}$ Also at State Key Laboratory of Nuclear Physics and Technology, Peking University, Beijing 100871, People's Republic of China\\
% ...
      }
    \vspace{0.4cm}
    \end{small}
  \end{center}
}

\begin{abstract}
Using data samples with a total integrated luminosity of $\lumtot$
at center-of-mass energies between $4.008$ and $4.951$~$\gev$ collected with the BESIII detector, the cross sections of $\ee\to\pizpiz\psip$ process are measured.
The obtained cross sections are found to be approximately one-half of those of $\ee\to\pipi\psip$, consistent with the isospin symmetry expectation.
A coherent fit to the dressed cross sections is performed, with the $\YI$~parameters fixed at the values measured in $\ee\to\pipi\psip$.
The significances of the $\YII$ and $\YIII$ are both larger than $5\sigma$, and their masses and widths are consistent with the previous measurement in the $\ee\to\pipi\psip$ process.

\end{abstract}

\maketitle

\section{INTRODUCTION}
Over the past two decades, a series of charmonium-like states with
$J^{PC}=1^{--}$, referred to as the $Y$ states, have been discovered
and confirmed by numerous experiments.
As exotic state candidates, the $Y$ states cannot be easily accommodated within
the conventional quark model~\cite{PRD72054026} and their internal properties remain
controversial.
Possible interpretations include hybrid mesons, meson molecules,
hadrocharmonia, and tetraquarks, but none of them has been proven
conclusive to date~\cite{PR6391121,RMP90015004,PR8731154,FPB21016300}.
Therefore, comprehensive studies of their production and decay patterns~\cite{arXiv250215117,PRD112054027},
as well as precise measurements of their resonant parameters from different
experiments, are highly desired to provide critical information to
explore the nature of the $Y$ states.

Among all $Y$ states, the $\YI$ and $\YII$, previously known as
$Y(4260)$ and $Y(4360)$, were first observed
by the {\it BABAR} and Belle experiments with the initial-state radiation (ISR) technique in
the processes of $\ee \to \gamma_{\rm ISR}\pipi\jpsi$~\cite{PRL95142001, PRL99182004, PRD86051102, PRL110252002}
and $\ee \to \gamma_{\rm ISR}\pipi\psip$~\cite{PRL98212001, PRD89111103, PRL99142002, PRD91112007}, respectively.
With high statistics data samples collected by the BESIII experiment,
the $Y(4260)$ and $Y(4360)$ were observed in several decay modes and measured
with improved precision.
Furthermore, it was found that $Y(4260)$ consists of two resonant structures,
$Y(4230)$ and $Y(4320)$~\cite{PRL118092001}.
The $\YI$ was also observed in the processes
$\ee \to \pizpiz\jpsi$~\cite{PRD102012009},
$\ee \to \ks\ks\jpsi$~\cite{PRD107092005},
$\ee \to \kp\km\jpsi$~\cite{CPC46111002},
$\ee \to \pipi\psip$~\cite{PRD96032004,PRD104052012},
$\ee \to \pipi\hc$~\cite{PRL118092002},
$\ee \to \omega\chicz$~\cite{PRL114092003,PRD99091103},
$\ee \to \pip D^{0}D^{*-}$~\cite{PRL122102002}, and
$\ee \to \eta\jpsi$~\cite{PRD102031101, PRD109092012}.
The $\YII$~was  observed in the processes
$\ee \to \pipi\hc$~\cite{PRL118092002},
$\ee \to \pipi\psi_{2}(3823)$~\cite{PRL129102003},
and $\ee \to \eta\jpsi$~\cite{PRD102031101, PRD109092012}.
Although there are many experimental studies of $\YI$ and $\YII$ with high precision,
the resonance parameters from various processes are different, which
needs further investigation.
Analogous to the $\YI$ and $\YII$, the $\YIII$ was first observed
by Belle and {\it BABAR} with the ISR process of $\ee \to \gamma_{\rm ISR}\pipi\psip$~\cite{PRL99142002, PRD91112007, PRD89111103},
and was also observed and confirmed by the BESIII experiment
in the processes $\ee \to \pipi\psip$~\cite{PRD104052012},
and $\ee \to \pipi\psi_{2}(3823)$~\cite{PRL129102003}.

The possible existence of $Y$ states in the $\pizpiz\psip$ final states
has been investigated by the BESIII experiment in Ref.~\cite{PRD97052001}.
However, no resonance-like structures were observed in the
cross section line shape due to the limited data statistics.
In this paper, an updated analysis of $\ee \to \pizpiz\psip$ at
center-of-mass (c.m.) energies ($\sqrt{s}$) between 4.008 and 4.951~GeV is performed.
The data samples used in this analysis were collected with the
BESIII detector at the BEPCII collider with an integrated luminosity
of $\lumtot$, including the data samples used in
the previous analysis~\cite{PRD97052001}.
The much larger data samples used in this analysis allow to
search for possible $Y$ states in the $\pizpiz\psip$ final state.
In this analysis, the $\psip$ is reconstructed via its decay to $\pipi\jpsi$,
the $\jpsi$ via $\jpsi \to \llep~(\ell=e/\mu)$ and the $\piz$ via $\piz \to \gamma\gamma$.

\section{Description of BEPCII and the BESIII detector}

The BESIII detector~\cite{NIMA614345399} records symmetric $e^+e^-$ collisions
provided by the BEPCII collider~\cite{7IPACIPAC20162016}
in the center-of-mass energy range from 1.84 to 4.95~GeV,
with a peak luminosity of $1.1 \times 10^{33}\;\text{cm}^{-2}\text{s}^{-1}$
achieved at $\sqrt{s} = 3.773\;\text{GeV}$.
BESIII has collected large data samples in this energy region~\cite{CPC44040001,RDTaM4337344,RDTaM6289293}.
The cylindrical core of the BESIII detector covers 93\% of the full solid angle and consists of a helium-based
multilayer drift chamber~(MDC), a time-of-flight
system~(TOF), and a CsI(Tl) electromagnetic calorimeter~(EMC),
which are all enclosed in a superconducting solenoidal magnet
providing a 1.0~T magnetic field.
The solenoid is supported by an
octagonal flux-return yoke with resistive plate counter muon
identification modules interleaved with steel.
The charged-particle momentum resolution at $1~{\rm GeV}/c$ is
$0.5\%$, and the ${\rm d}E/{\rm d}x$ resolution is $6\%$ for electrons
from Bhabha scattering. The EMC measures photon energies with a
resolution of $2.5\%$ ($5\%$) at $1$~GeV in the barrel (end cap)
region. The time resolution in the plastic scintillator TOF barrel region is 68~ps, while
that in the end cap region was 110~ps. The end cap TOF
system was upgraded in 2015 using multigap resistive plate chamber
technology, providing a time resolution of 60~ps~\citep{RDTM1,RDTM115, NIaMiPRSAASDaAE953163053}, which benefits 76.5\% of
the data used in this analysis.

Monte Carlo (MC) simulated data samples produced with a {\sc
geant4}-based~\cite{NIMA506250303} software package, which
includes the geometric description of the BESIII detector and the
detector response, are used to determine detection efficiencies
and to estimate backgrounds. The simulation models the beam
energy spread and ISR in the $e^+e^-$
annihilation with the generator {\sc kkmc}~\cite{PRD63113009}.
The inclusive MC sample includes the production of open charm
processes, the ISR production of vector charmonium(-like) states,
and the continuum processes incorporated in {\sc kkmc}~\cite{PRD63113009}.
All particle decays are modelled with {\sc
evtgen}~\cite{NIMA462152155, CPC32599} using branching fractions
either taken from the Particle Data Group (PDG)~\cite{PRD110030001} when available,
or otherwise estimated with {\sc lundcharm}~\cite{PRD62034003, CPL31061301}.
Final state radiation~(FSR) from charged final state particles is incorporated using the
{\sc photos} package~\cite{CPC66115128}.
Signal MC samples of $\ee\to\pizpiz\psip$ are generated
with the amplitude model from the partial wave analysis (PWA) in Ref.~\cite{PRD104052012}.

\section{EVENT SELECTION}

The event selection criteria are designed to ensure reliable identification of the final-state particles and proper reconstruction of the event.
Good charged tracks detected in the MDC are required to be within
a polar angle ($\theta$) range of $|\rm{cos\theta}|<0.93$,
where $\theta$ is defined with respect to the $z$-axis,
which is the symmetry axis of the MDC.
The distance of closest approach to the interaction point (IP)
must be less than 10\,cm along the $z$-axis $|V_{z}|$,
and less than 1\,cm in the plane perpendicular to $z$-axis $|V_{xy}|$.
Photon candidates are identified using isolated showers in the EMC.
The deposited energy of each shower must be more than 25~$\mev$ in the barrel
region ($|\cos \theta|< 0.80$) or more than 50~$\mev$ in the end cap
region ($0.86 <|\cos \theta|< 0.92$).
To exclude showers that originate from charged tracks,
the angle subtended by the EMC shower and the position of
the closest charged track at the EMC
must be greater than 10 degrees as measured from the IP.
To suppress electronic noise and showers unrelated to the collision event,
the difference between the EMC time and the event start time
is required to be within [0, 700]~ns.

The detailed MC study indicates the pions and leptons
of signal are well-separated kinematically. Therefore the charged tracks
with momenta above 1.1~$\gevc$ are assigned to be leptons,
while those with momenta below 0.75~$\gevc$ are pions.
Electron and muon are separated by using the deposited energy ($E$)
in the EMC, i.e., both electron and positron must satisfy $E/p > 0.7$,
while both muons must satisfy $E < 0.45$~$\gev$, where $p$ is the momentum of charged
tracks measured by the MDC. Signal candidates are required to have a pair of charged
pions with zero net charge and a lepton pair with the same flavor
and opposite charge.

To improve the resolution and suppress the background,
a four-constraint (4C) kinematic fit imposing energy-momentum conservation
under the hypothesis of $\ee\to\gam\gam\gam\gam\pipi\llep$ is
performed, and $\chi_{4\rm C}^{2}<120$ is required.
For events with more than four photons, the combination
with the smallest $\chi_{4\rm C}^{2}$ is retained.
The four selected photons are combined into two $\piz$ candidates
by minimizing
$(M(\gam_{1}\gam_{2})-M_{\piz})^2+(M(\gam_{3}\gam_{4})-M_{\piz})^2$,
where $M(\gam_{i}\gam_{j})$ is the invariant mass of two
photons and $M_{\piz}$ is the $\piz$ nominal mass quoted from
the PDG~\cite{PRD110030001}.
Both $\piz$ candidates are further required to satisfy
$|M(\gam_{i}\gam_{j})-M_{\piz}|<20$~$\mevcsq$, which is
larger than five times the $\piz$ mass resolution.
In order to improve the mass resolution, a seven-constraint (7C)
kinematic fit with the additional mass constraints on the two $\piz$'s and
$\jpsi$ candidates to their nominal masses is carried out.
The particle kinematics information updated by the 7C kinematic fit is used in the following analysis, and the $\psi(3686)$ candidate is reconstructed from the $\llep$ pair and
charged pion pair.

\section{EXTRACTION OF THE BORN CROSS SECTIONS}

\begin{figure*}[htbp]
	\centering
{
		\begin{overpic}[width=0.2\linewidth]{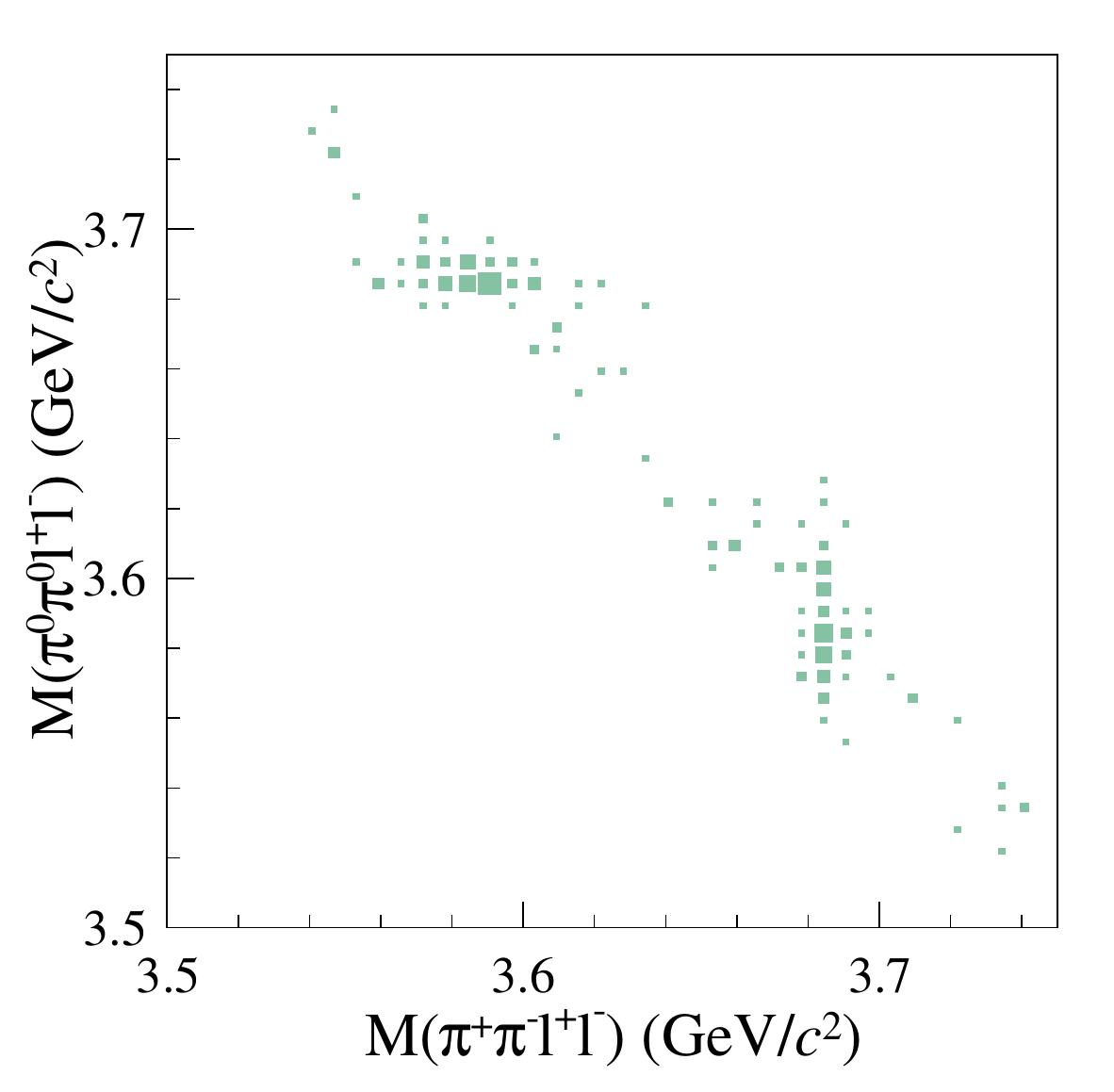}
			\put(82,82){\scriptsize (a)}
		\end{overpic}
	}
	\hspace{-1.35em}
{
		\begin{overpic}[width=0.2\linewidth]{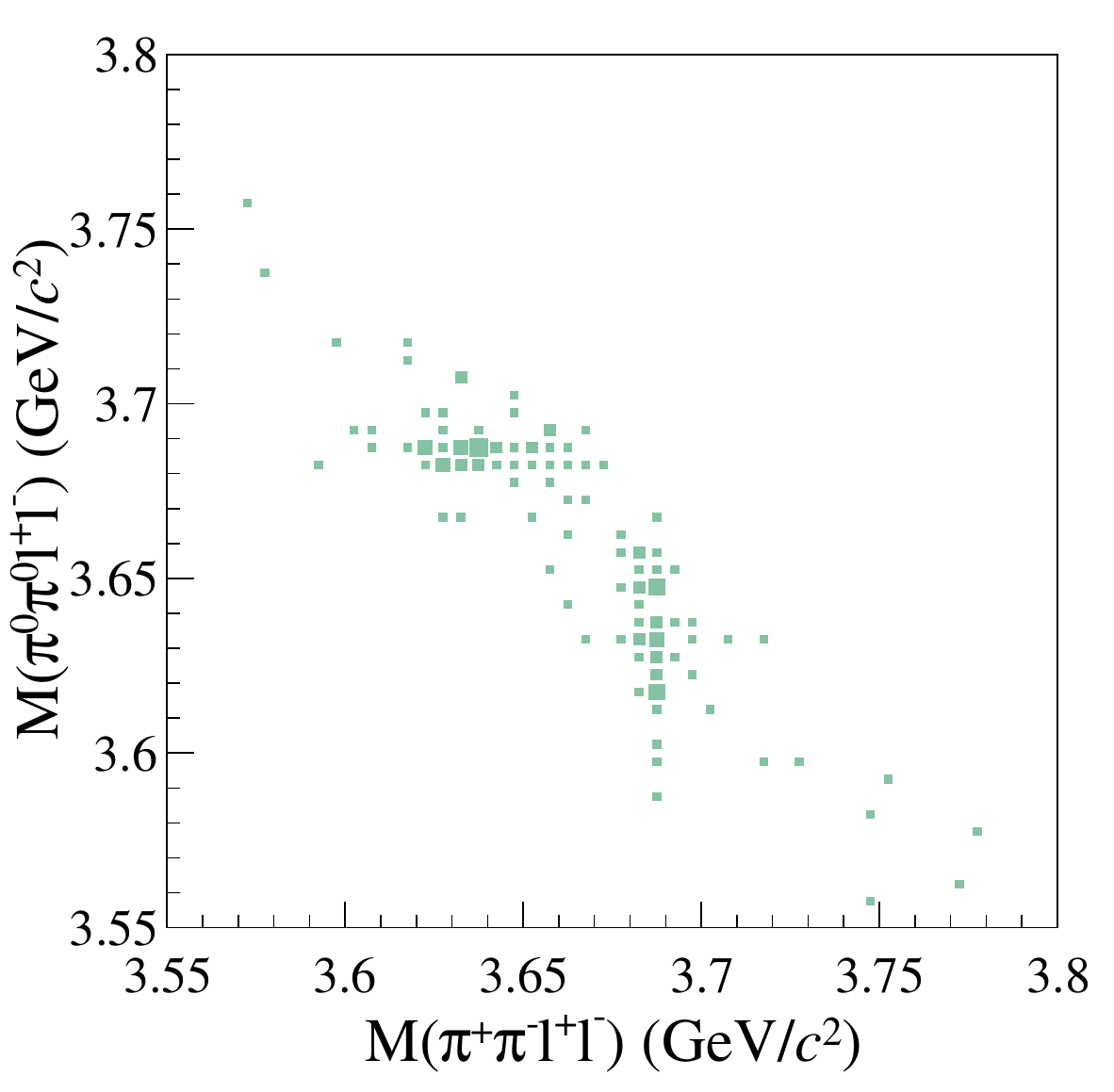}
			\put(82,82){\scriptsize (b)}
		\end{overpic}
	}
	\hspace{-1.35em}
{
		\begin{overpic}[width=0.2\linewidth]{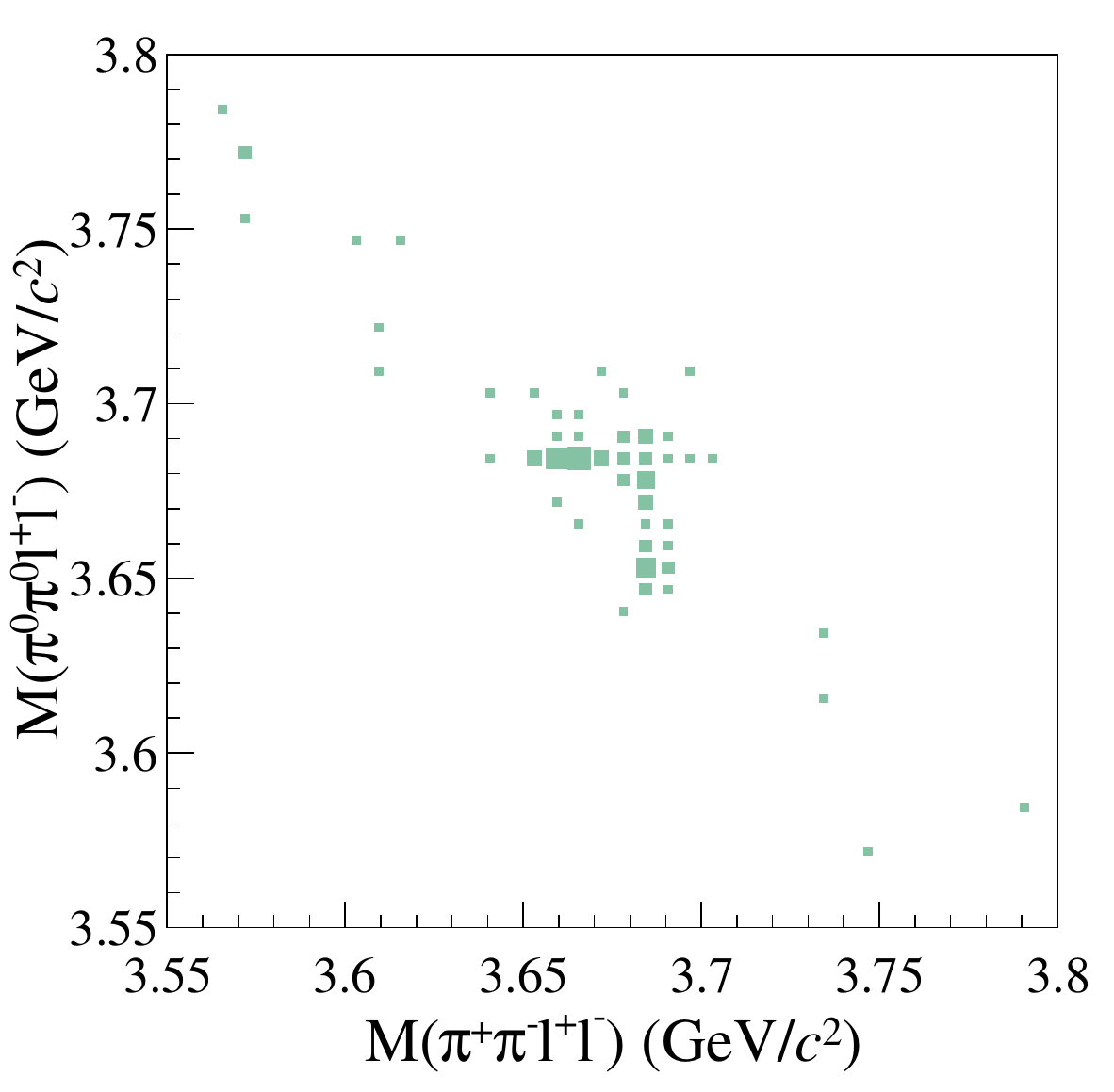}
			\put(82,82){\scriptsize (c)}
		\end{overpic}
	}
	\hspace{-1.35em}
{
		\begin{overpic}[width=0.2\linewidth]{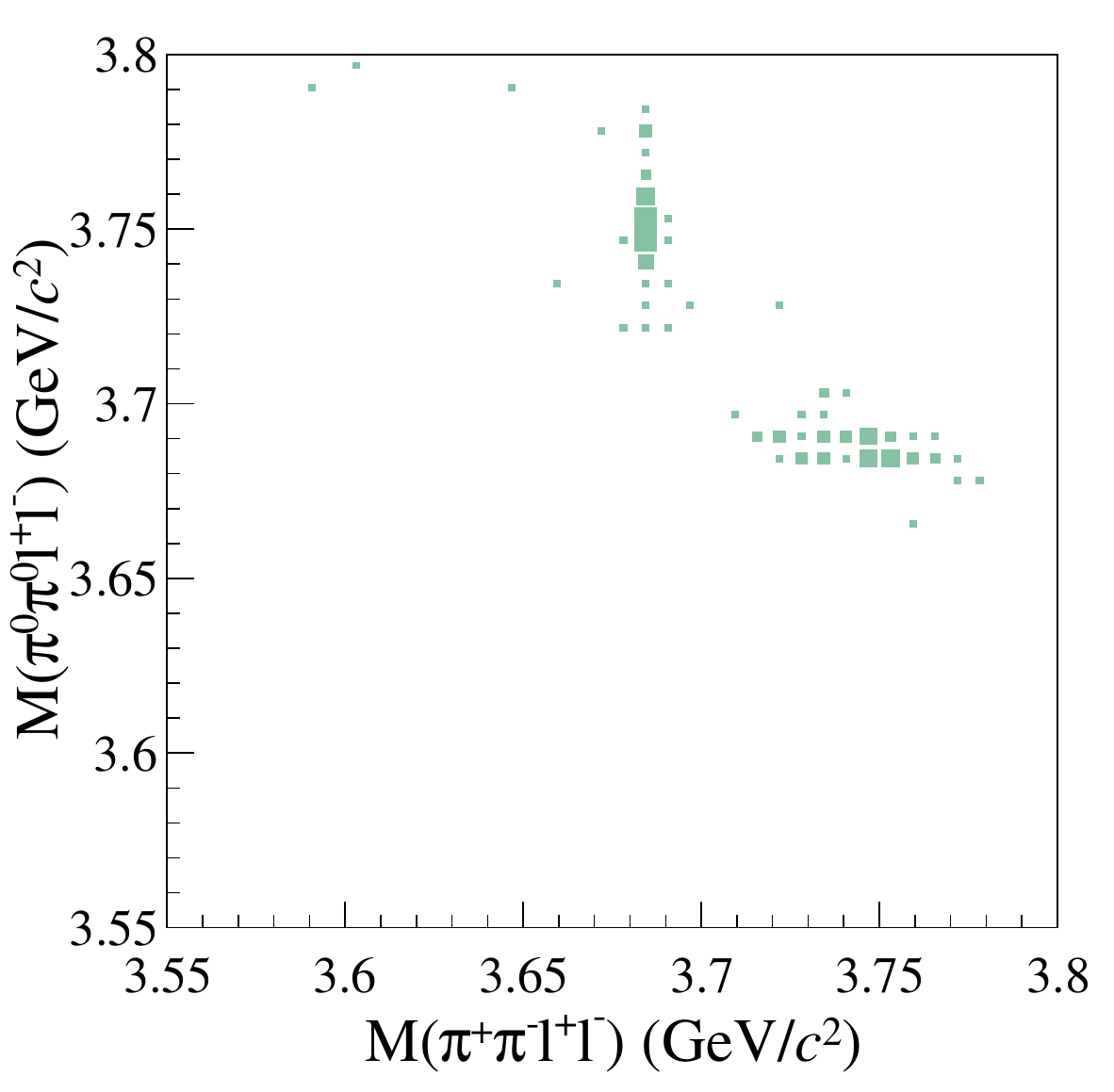}
			\put(82,82){\scriptsize (d)}
		\end{overpic}
	}
	\hspace{-1.35em}
{
		\begin{overpic}[width=0.2\linewidth]{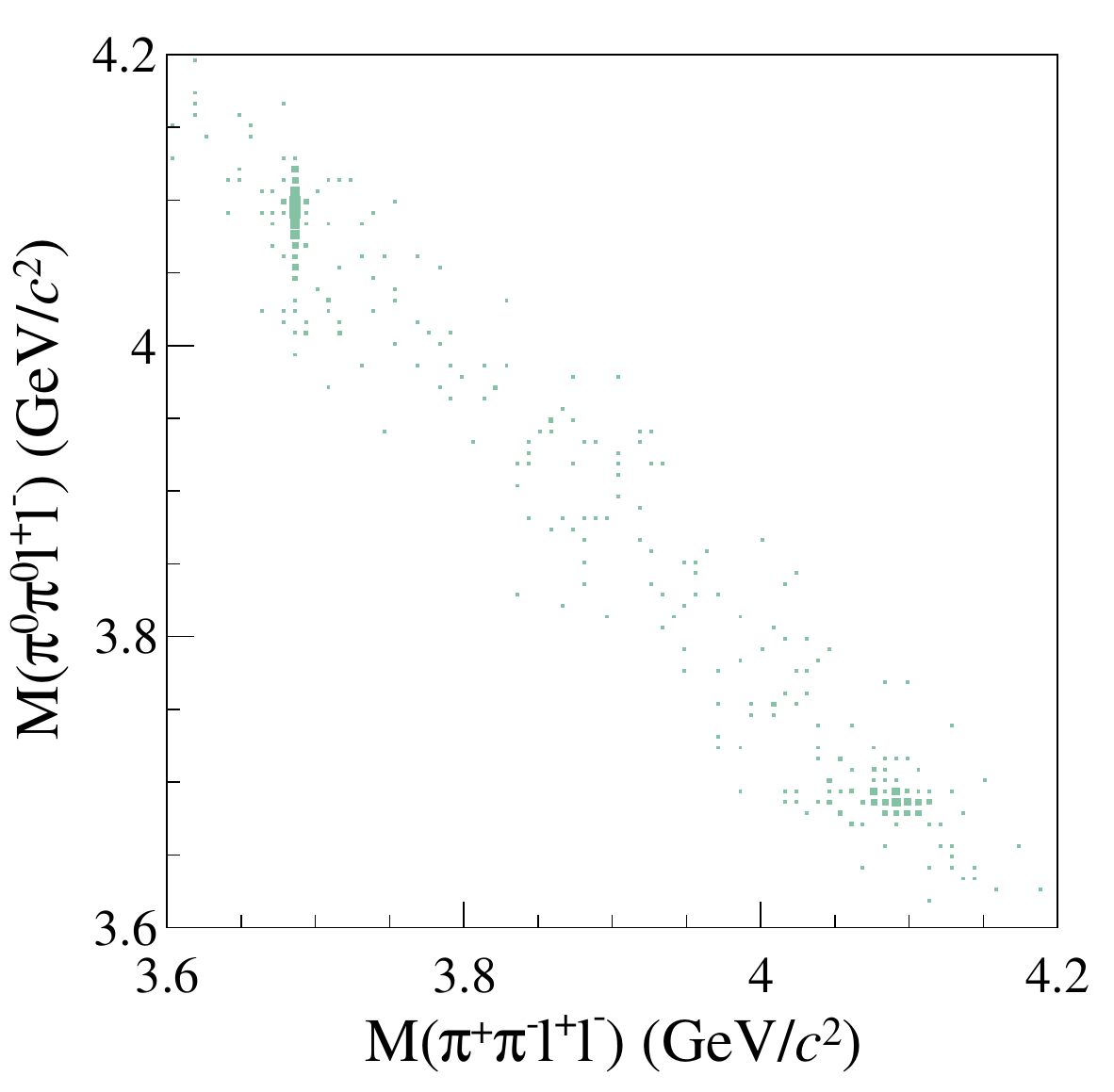}
			\put(82,82){\scriptsize (e)}
		\end{overpic}
	}

	\caption{Distributions of $M(\pipi\llep)$ versus $M(\pizpiz\llep)$
    for high statistics data sets taken at $\sqrt{s}=$
  (a) 4.1784, (b) 4.2263, (c) 4.2580,
  (d) 4.3374, and (e) 4.6819 \gev, respectively. }
  \label{fig:2Dband_data}
\end{figure*}

From the studies of inclusive and exclusive MC samples,  it is found that
the main background is from the process
$\ee\to\pipi\psip$ with the subsequent decay $\psip\to\pizpiz\jpsi$, which has the same final state as the signal.
As shown in the two-dimensional distributions of
$M(\pipi\llep)$ versus $M(\pizpiz\llep)$ for data
at c.m. energies with high statistics in Fig.~\ref{fig:2Dband_data},
there are two bands corresponding to the
signal (the vertical one) and background (the horizontal one)
events, respectively.
For data sets around $\sqrt{s}=4.26~\gev$,
the signal and background bands partially overlap due to the similar kinematic distributions of $\pip\pim$  and $\piz\piz$ pairs.
Therefore, the background events are included in the fit when extracting
signal yields for those data sets in the $\sqrt{s}$ range from $4.1989~\gev$
to $4.3583~\gev$.

\begin{figure*}[htbp]
	\centering
{
		\begin{overpic}[width=0.2\linewidth]{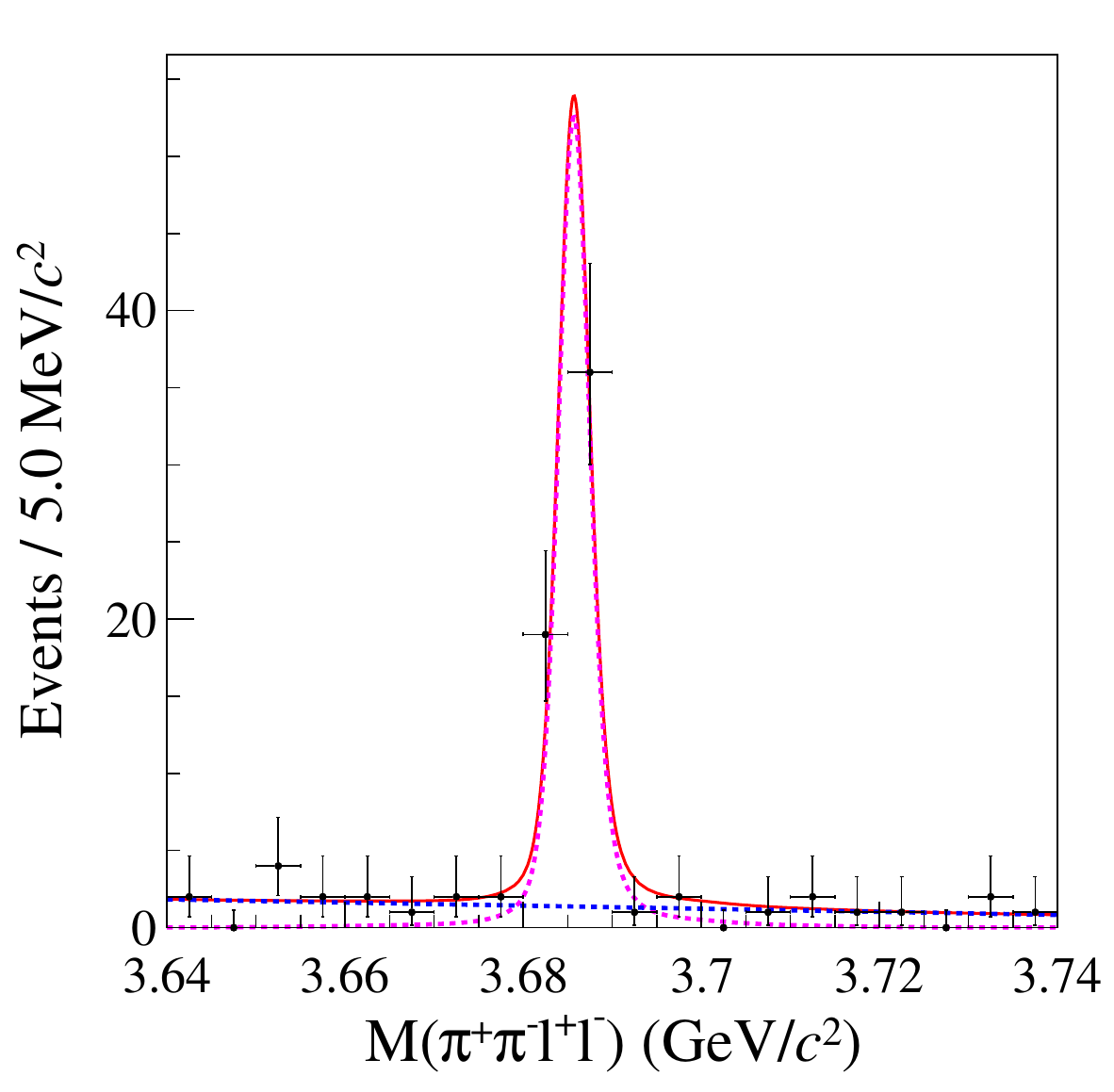}
			\put(82,82){\scriptsize (a)}
		\end{overpic}
	}
	\hspace{-1.35em}
{
		\begin{overpic}[width=0.2\linewidth]{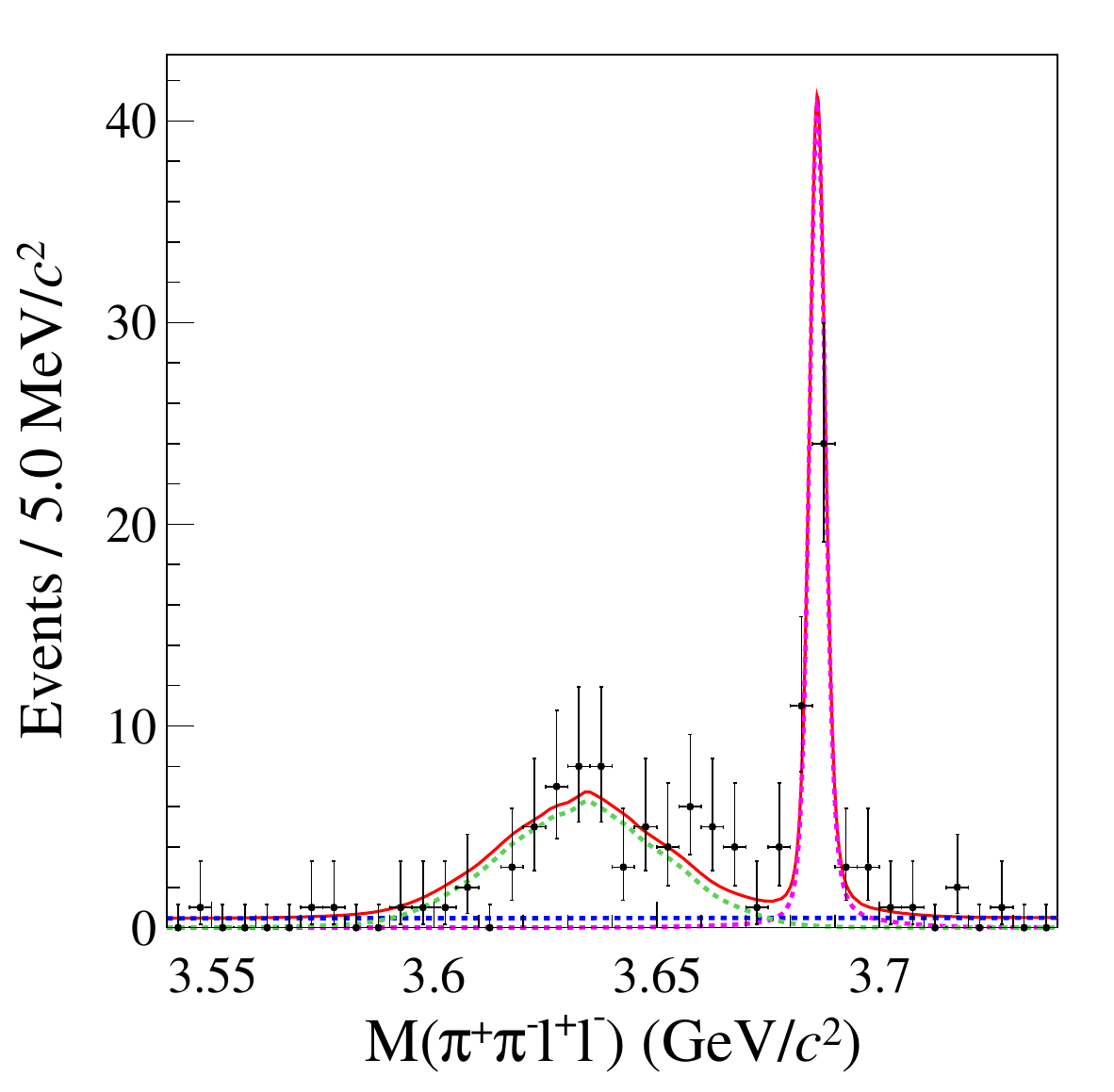}
			\put(82, 82){\scriptsize (b)}
		\end{overpic}
	}
	\hspace{-1.35em}
{
		\begin{overpic}[width=0.2\linewidth]{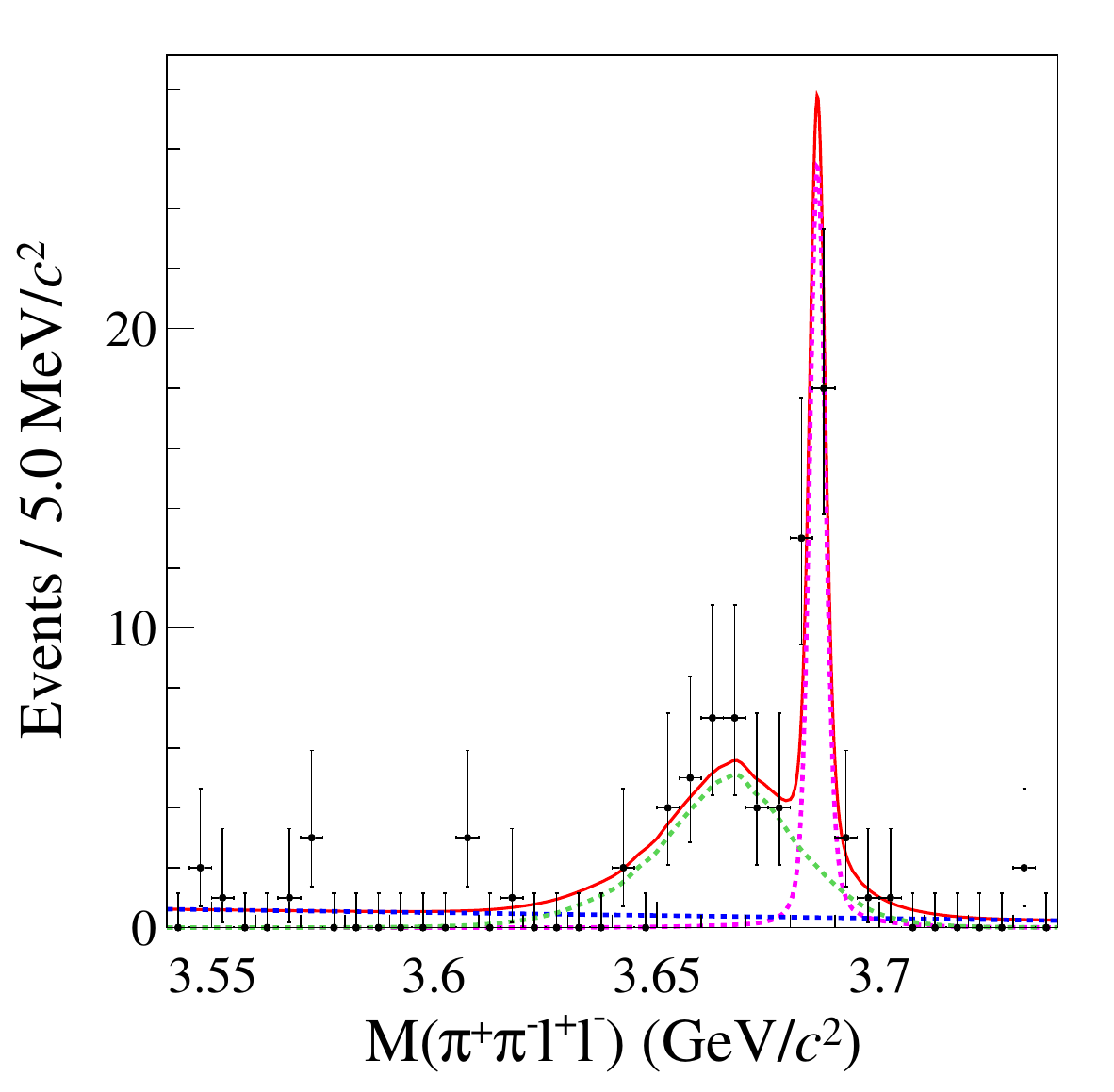}
			\put(82, 82){\scriptsize (c)}
		\end{overpic}
	}
	\hspace{-1.35em}
{
		\begin{overpic}[width=0.2\linewidth]{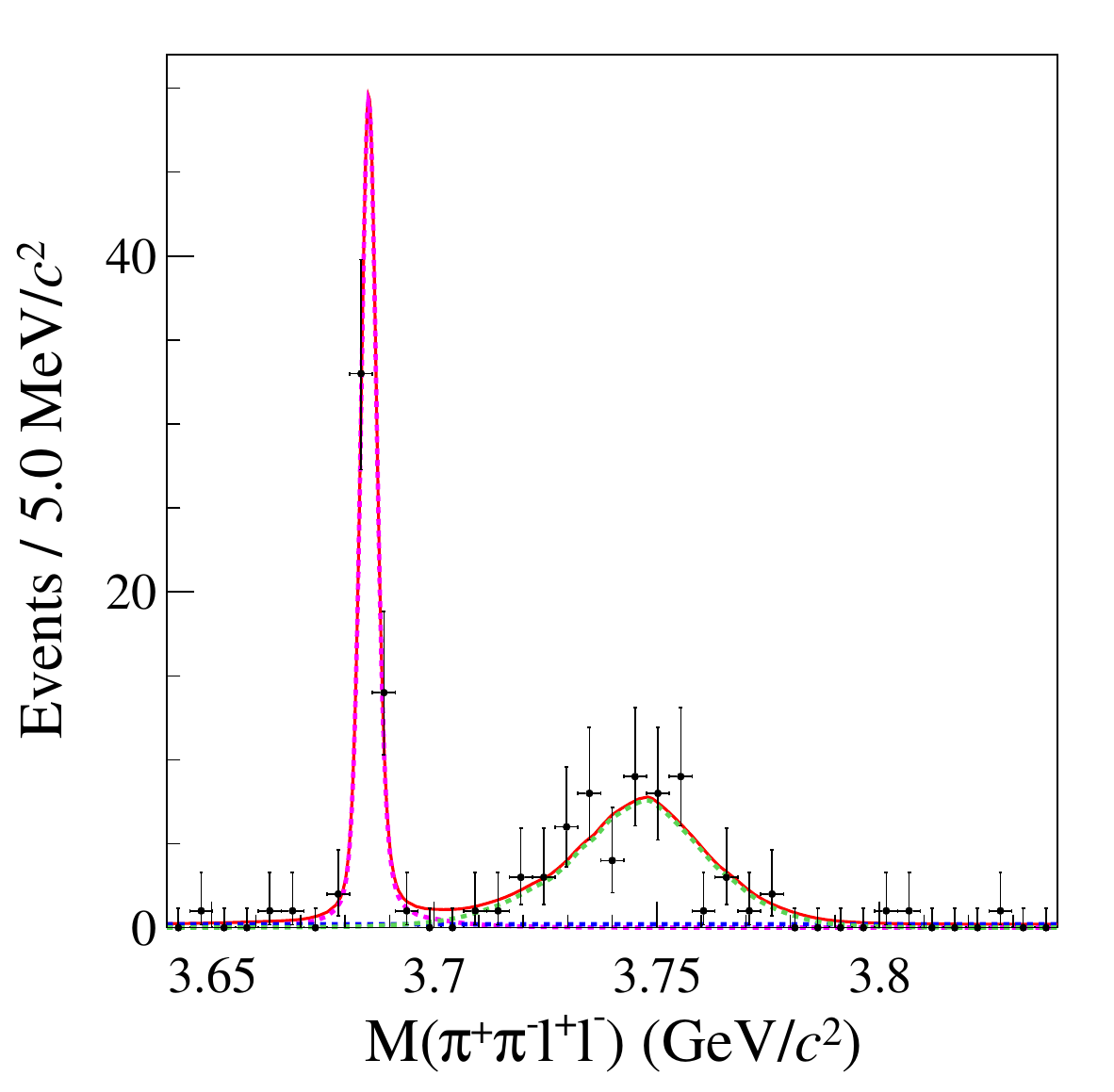}
			\put(82, 82){\scriptsize (d)}
		\end{overpic}
	}
	\hspace{-1.35em}
{
		\begin{overpic}[width=0.2\linewidth]{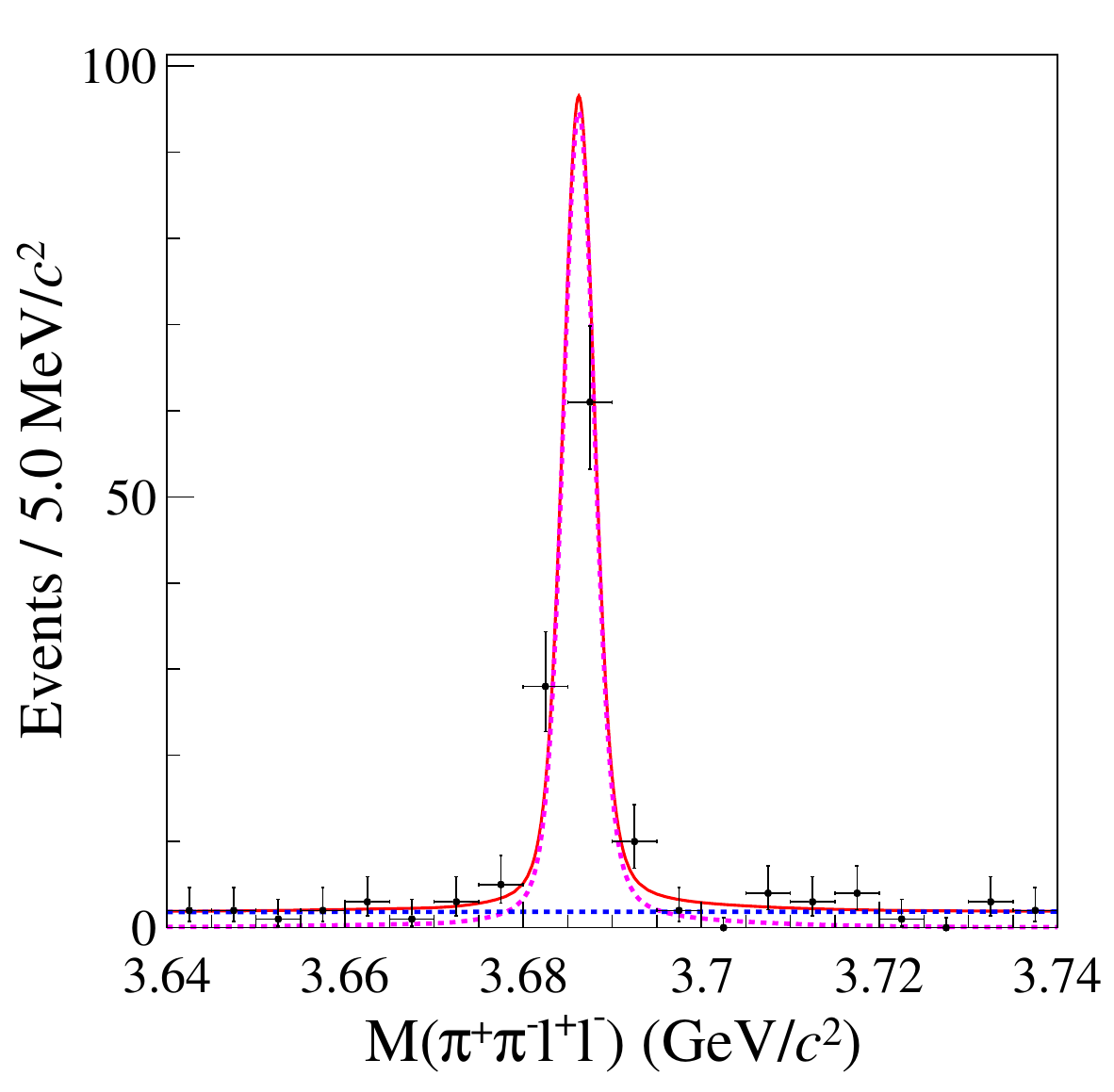}
			\put(82, 82){\scriptsize (e)}
		\end{overpic}
	}

	\caption{
  Fits to the $M(\pipi\llep)$ distributions
  for high statistics data sets taken at $\sqrt{s}=$
  (a) 4.1784, (b) 4.2263, (c) 4.2580,
  (d) 4.3374, and (e) 4.6819 $\gev$, respectively.
  The black dots with error bars are data, the red solid lines are
  the fit results, the dashed-red lines are the signal components,
  the dashed-green lines are the background components, and the
  dashed-blue lines are the non-peaking background components. }
  \label{fig:fit_result}
\end{figure*}
\begin{figure}[htbp]
	\centering
    \includegraphics[scale=0.45]{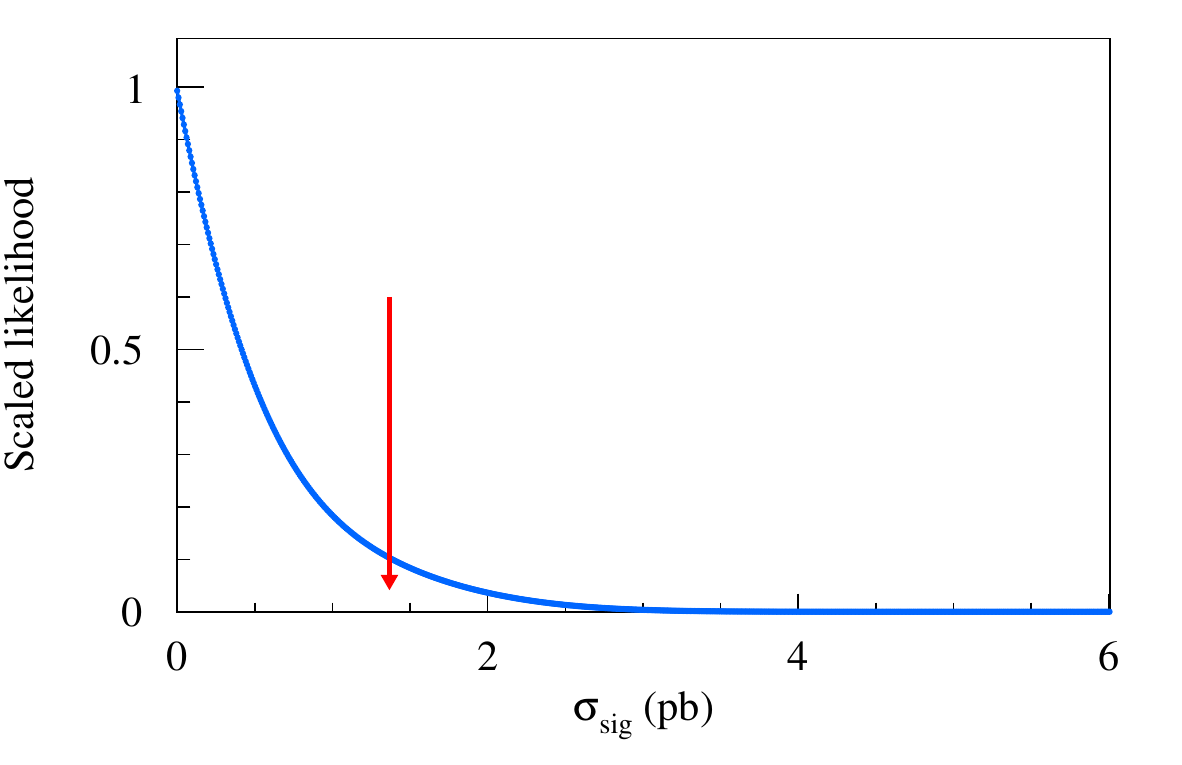}
	\caption{
    The scaled likelihood distribution, normalized to its maximum value, as a function of the cross section
    at $\sqrt{s}=4.0076$ GeV. The red arrow indicates the upper limit
    at the 90\% confidence level.
  }
  \label{fig:upper_limit}
\end{figure}

The signal yields at the different c.m. energies are extracted by performing unbinned maximum
likelihood fits on the $M(\pipi\llep)$ distribution.
In these fits, the fit ranges are chosen according to the background situation.
The signal shapes are described
with the exclusive MC-simulated shape convolved with a Gaussian function,
which accounts for the resolution difference between data and MC simulation.
Among the different data sets, the widths of the Gaussian functions
are all fixed to the values obtained from simultaneous fits for
four high statistics data sets taken at $\sqrt{s}$=4.1784, 4.3774, 4.4156, 4.6819~$\gev$.
The shapes for the main background process
$\ee\to\pipi\psip$ with the subsequent decay $\psip\to\pizpiz\jpsi$ and
other non-peaking background are described by
the MC-simulated shapes and first-order polynomial functions, respectively. The fits to the $M(\pipi\llep)$ distributions for high statistics data sets are shown in Fig.~\ref{fig:fit_result}.

The Born cross section is determined by the formula
\begin{equation}
  \sigma^{\mathrm{B}}=\frac{N_{\mathrm{sig}}}{\mathcal{L}_{\mathrm{int}} \cdot\left(1+\delta^{\mathrm{ISR}}\right) \cdot \frac{1}{|1-\Pi|^2} \cdot \mathcal{B} \cdot \epsilon},
  \label{eq:xsec}
\end{equation}
where $N_{\mathrm{sig}}$ is the number of signal events obtained
from the fit to the $M(\pipi\llep)$ distribution,
$\mathcal{L}_{\mathrm{int}}$ is the integrated luminosity,
$1+\delta^{\mathrm{ISR}}$ is the ISR correction factor,
$\frac{1}{|1-\Pi|^2}$ is the vacuum polarization
correction factor taken from Ref.~\cite{EPJC66585686},
$\mathcal{B}$ is the branching fraction of the subsequent decay quoted from the PDG~\cite{PRD110030001},
and $\epsilon$ is the detection efficiency.
The ISR correction factor and the detection efficiency
are estimated using the signal MC samples. An iterative
weight method~\cite{FPB1664501} is used to consider the influence
of the cross-section line shape and to estimate the ISR
correction factor. The relationship between the dressed cross section
and the Born cross section is given by
$\sigma^{\mathrm{D}}=\sigma^{\mathrm{B}}/|1-\Pi|^2$.

For those low statistics data sets with
the signal significance less than 3$\sigma$, the
cross section upper limits are estimated using the
Bayesian method~\cite{RoPiP661383}.
As an example, the likelihood distribution at $\sqrt{s}=4.0076$~$\gev$ is shown
in Fig.~\ref{fig:upper_limit}.
With a uniform prior probability density function, the
Bayesian upper limit at 90\% confidence level (C.L.)
is estimated by integrating the likelihood distribution
from zero to the upper limit, which gives 90\% of the
total area.

\begin{figure}[htbp]
	\centering
    \includegraphics[scale=0.45]{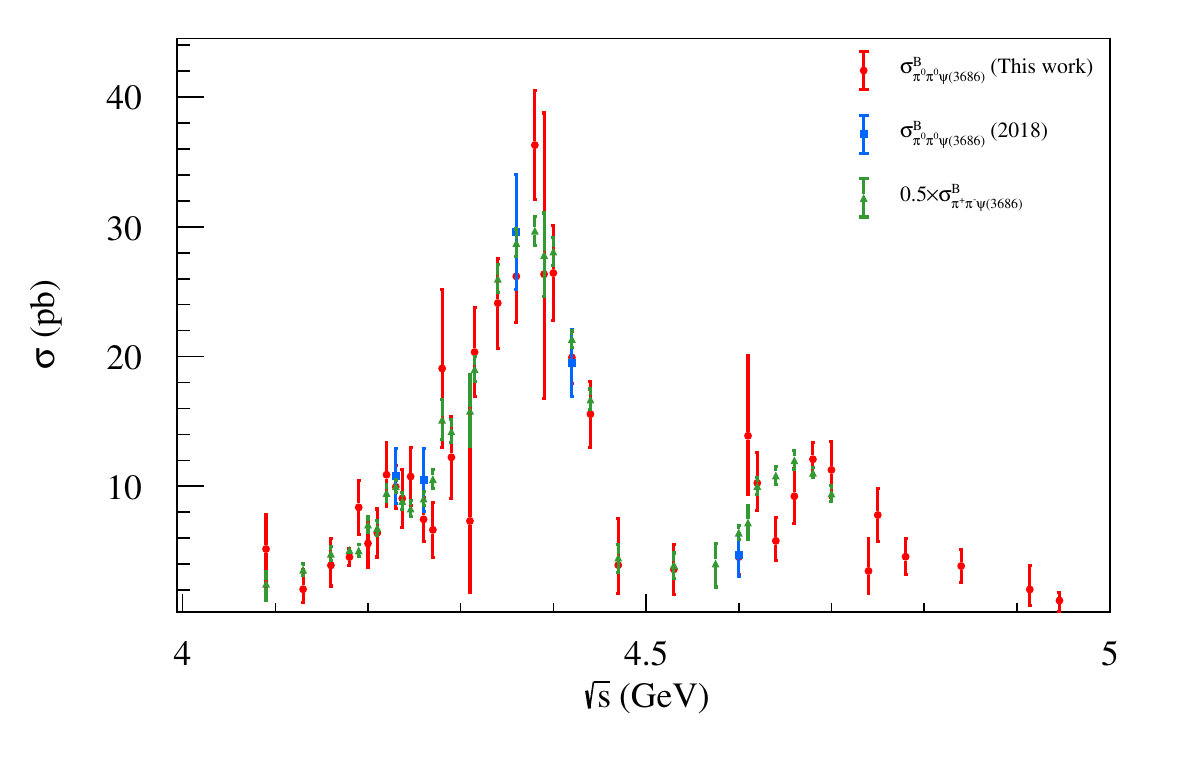}
	\caption{
    The comparison of the cross-section line shapes.
    The red dots are results from this work;
    the blue squares are results of $\ee\to\pizpiz\psip$ from Ref.~\cite{PRD97052001};
    the green triangles are results of $\ee\to\pipi\psip$ from Ref.~\cite{PRD104052012}.
    The cross sections of $\ee\to\pipi\psip$ have been multiplied by 0.5
    for better comparison between the two processes.}
  \label{fig:xsec_comp}
\end{figure}

The Born cross sections for each c.m.~energy
and other related quantities used in the Born cross section calculation are summarized in
Table~\ref{tab:xsec_result}.
A comparison of the cross section line shapes between  $\ee\to\pizpiz\psip$ from this work and previous work~\cite{PRD97052001} and that from $\ee\to\pipi\psip$~\cite{PRD104052012}
is shown in Fig.~\ref{fig:xsec_comp}.
Similar resonant structures are observed in the cross-section line shapes of both processes.

\section{FIT TO THE DRESSED CROSS SECTIONS}
To study the possible resonant structures in the process
$\ee\to\pizpiz\psip$, an extended maximum likelihood fit
is performed to the dressed cross sections at different c.m. energies. For data sets
with high statistics in Table~\ref{tab:xsec_result},
the cross section is expected to
follow an asymmetric Gaussian distribution $G_{i}$.
Following Ref.~\cite{PRL118092002},
$G_i$ is defined as
\begin{equation}
  \begin{aligned}
    & G_i(\sigma_{i}^{\text{fit}}(\sqrt{s}))=\frac{1}{\delta_i \sqrt{2 \pi}} e^{-\frac{\left(\sigma_{i}^{\text{fit}}(\sqrt{s})-\sigma_{i}\right)^2}{2 \delta_i^2}}, \\
    & \delta_i=
    \begin{cases}
    \delta_i^{\text {negative}} & \sigma^{\text{fit}}(\sqrt{s})<\sigma_i \\ \delta_i^{\text {positive}} & \sigma^{\text{fit}}(\sqrt{s}) \geq \sigma_i
    \end{cases},
  \end{aligned}
  \label{eq:xsec_pdf}
\end{equation}
where $\sigma_{i}^{\text{fit}}(\sqrt{s})$ is the theoretical cross section
as defined in Eq.~\ref{eq:fitmodel},
$\sigma_{i}$ is the measured cross section of the $i$-th data point,
and $\delta_i$ is the statistical uncertainty, which is set to
 $\delta_i^{\text{negative}}$ and $\delta_i^{\text{positive}}$
for the cross section below and above the measured value, respectively.
For the low statistics data sets, the corresponding cross section likelihood
distributions, which are parameterized as $L(\sigma^{\rm fit}_{i}(\sqrt{s}))$
with double-Gaussian function as shown in Fig.~\ref{fig:upper_limit}, are
used to describe the cross section distributions in the fit. The total likelihood function is defined as the product of
the likelihood functions of all data sets, which is
\begin{equation}
  L=\prod_{i=1}^{N_{1}=31} G_i(\sigma^{\text{fit}}) \cdot \prod_{j=1}^{N_{2}=10} L_j(\sigma^{\text{fit}})
\end{equation}
, where $N_{1}$ and $N_{2}$ are the numbers of data points
with or without sufficient statistics, respectively.

The cross-section line shape of $\ee\to\pizpiz\psip$ is consistent
with that of $\ee\to\pipi\psip$~\cite{PRD104052012} and
three resonant structures are observed there.
Therefore, a model including the coherent sum of three resonant structures
$\YI$, $\YII$ and $\YIII$ as well as the continuum is used to describe the cross section:
\begin{equation}
	\sigma^{\mathrm{fit}}(\sqrt{s})=\left|\sum_k e^{i \phi_k} \cdot B W_k(s)+e^{i \phi_{\mathrm{cont}}} \cdot \psi_{\mathrm{cont}}\right|^2,
	\label{eq:fitmodel}
\end{equation}
where $BW_{k}$ is the Breit-Wigner function for the $k$-th resonance,
the $\psi_{\mathrm{cont}}$ is the non-resonant function, and
the $\phi_{k}$ and $\phi_{\mathrm{cont}}$ are the corresponding phase
relative to the $Y(4390)$.
The $BW_{k}$ is defined as
\begin{equation}
  BW_k(s)=\frac{M_k}{\sqrt{s}} \frac{\sqrt{12 \pi \Gamma_k^{\mathrm{tot}} \Gamma_k^{e e} {\mathcal B}_k}}{s-M_k^2+i M_k \Gamma_k^{\mathrm{tot}}} \sqrt{\frac{\Phi(\sqrt{s})}{\Phi\left(M_k\right)}},
\end{equation}
where $M_k$, $\Gamma_k^{\mathrm{tot}}$, $\Gamma_k^{e e}$,
and ${\mathcal B}_k$ are the mass, total width, the partial width
coupling to $\ee$, and the branching fraction of the $k$-th
resonance, respectively. The $\Phi(\sqrt{s})$ is the standard
phase space factor~\cite{PRD110030001}.
The continuum part is parameterized as
\begin{equation}
	\psi_{\mathrm{cont}}=\frac{a}{(\sqrt{s})^n} \sqrt{\Phi(\sqrt{s})}
\end{equation}
with the free parameters $a$ and $n$.

\begin{figure*}
	{
		\begin{overpic}[width=0.45\linewidth]{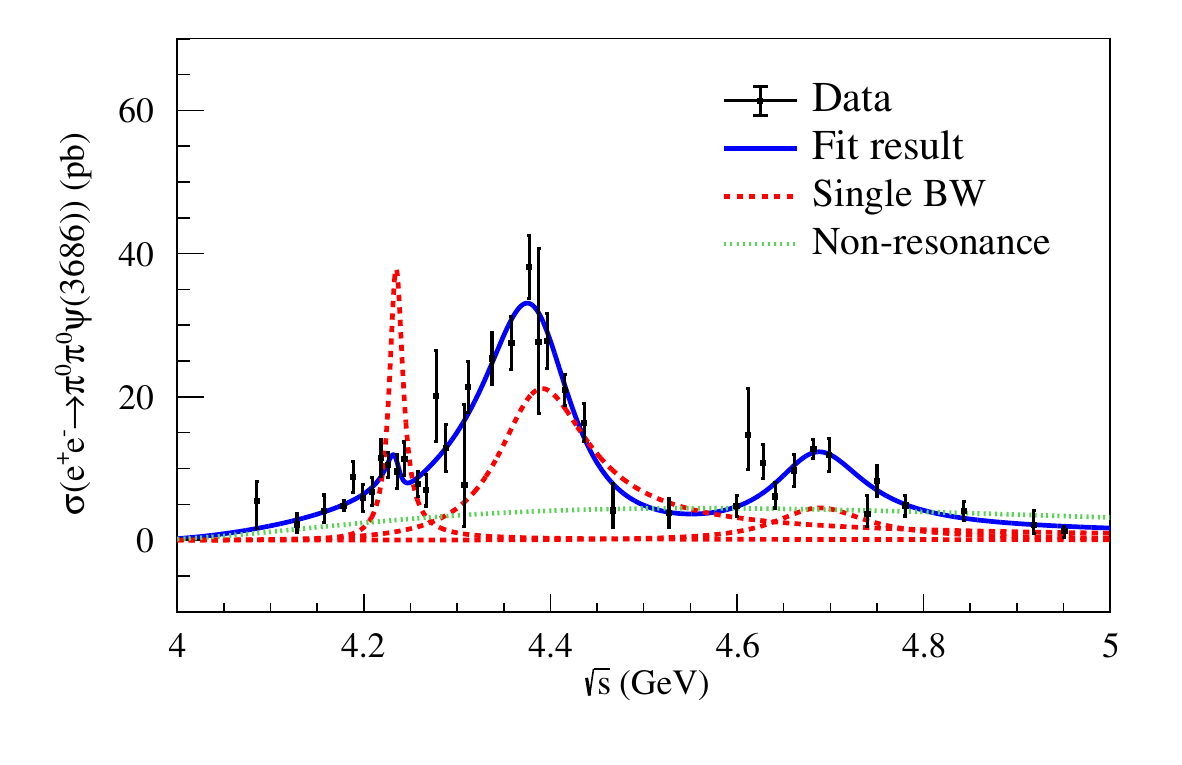}
			\put(20, 52){\small Solution I}
		\end{overpic}
	}
	{
		\begin{overpic}[width=0.45\linewidth]{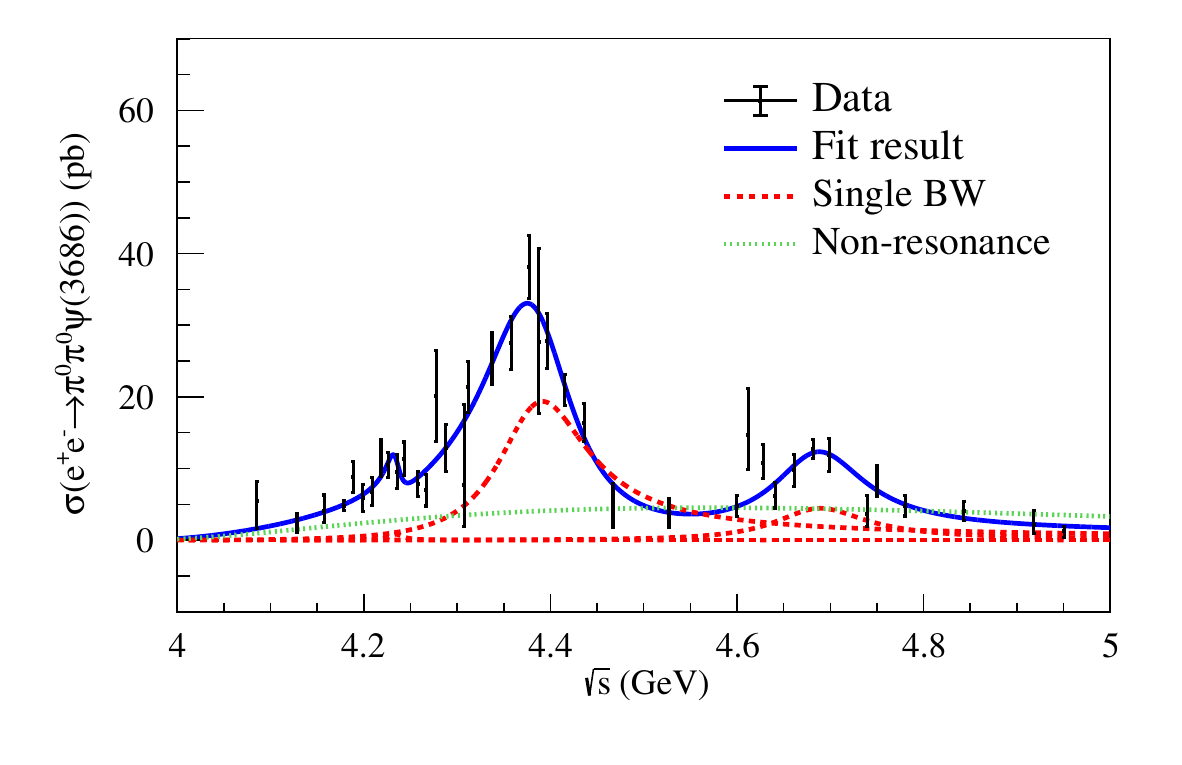}
			\put(20, 52){\small Solution II}
		\end{overpic}
	}

	{
		\begin{overpic}[width=0.45\linewidth]{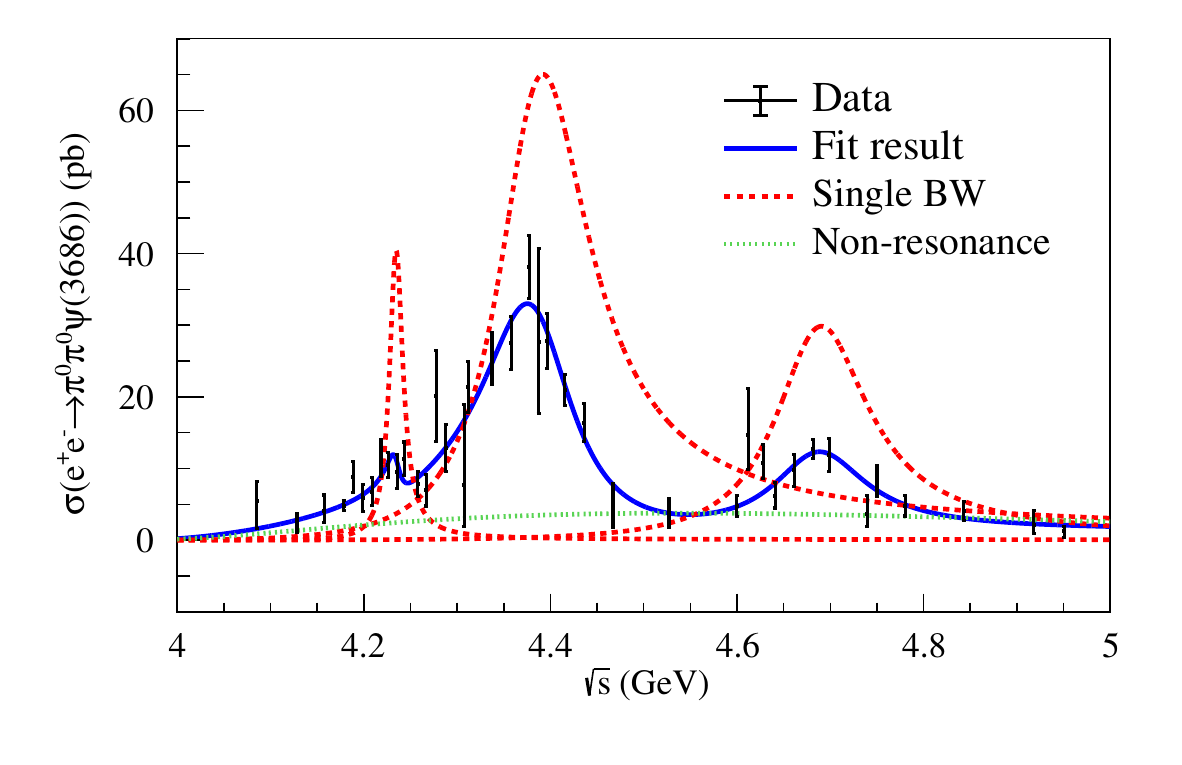}
			\put(20, 52){\small Solution III}
		\end{overpic}
	}
	{
		\begin{overpic}[width=0.45\linewidth]{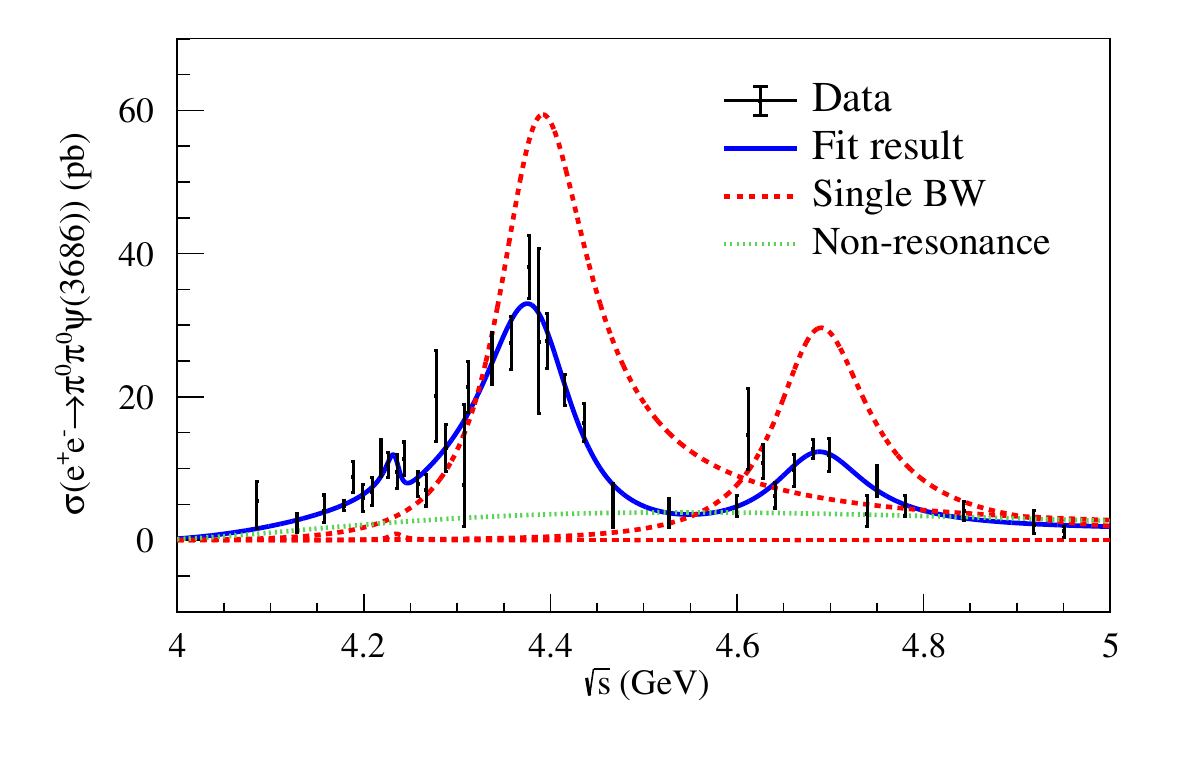}
			\put(20, 52){\small Solution IV}
		\end{overpic}
	}
	\caption{
		Fits to dressed cross sections of
		$\ee\to\pizpiz\psip$ with four solutions
		as summarized in Table~\ref{tab:sol_xsec_fit}.
		The black dots with error bars are the measured dressed cross sections,
		the blue-solid curves are the best fit results,
		the red-dashed lines represent individual resonant structures,
		and the green-dashed lines are the continuum components.
}\label{fig:sol_xsec_fit}
\end{figure*}

With the above model, the fit to the
dressed cross sections is performed by maximizing the
likelihood function $L$.
The statistical significance of the
resonant structures is estimated by comparing the
likelihood values of the fits with and without the
corresponding resonant structures included.
The resultant significances of the resonant structures are
$1.9\sigma$, $11.4\sigma$, and $5.7\sigma$ for
$\YI$, $\YII$, and $\YIII$, respectively.
Since the significance of $\YI$ is not substantial,
the parameters of $\YI$ are fixed to
the values from the measurement of $\ee\to\pipi\psip$~\cite{PRD104052012},
while the parameters of $\YII$ and $\YIII$ are
free in the fit.
By scanning three relative phases, four solutions are
found with similar fit quality and identical
masses and widths of $\YII$ and $\YIII$.
The fit results are shown in Table~\ref{tab:sol_xsec_fit}
and Fig.~\ref{fig:sol_xsec_fit}.

\begin{table*}[htbp]
	\centering
	\caption{
   For each c.m. energy $\sqrt{s}$, the following quantities are listed:
   the integrated luminosity $\mathcal{L}_{\mathrm{int}}$,
   the ISR correction factor $1+\delta^{\mathrm{ISR}}$,
   the vacuum polarization correction factor $\frac{1}{|1-\Pi|^2}$,
   the selection efficiency $\epsilon$,
   the signal yield $N_{\mathrm{sig}}$,
   the Born cross section $\sigma^{\mathrm{B}}$,
   the ratio of the cross sections between $\ee\to\pizpiz\psip$ and
   that of $\ee\to\pipi\psip$ denoted as $R$,
   the cross section upper limit $\sigma^{\mathrm{UP}}$ at the 90\% C.L.
   The first uncertainties are statistical and the second systematic.
  }
	\label{tab:xsec_result}
	\scalebox{1.00}{
		\begin{tabular}{c|c|c|c|c|D{,}{}{10.7}|D{,}{}{9.7}|D{,}{}{8.7}|c}
			\hline
			\hline
      $\sqrt{s}~(\gev)$ & $\mathcal{L}_\mathrm{int}~$(pb$^{-1}$)   & $1+\delta^{\mathrm{ISR}}$ & $\frac{1}{|1-\Pi|^2}$ & $\epsilon~$(\%) & N_{\mathrm{sig}}, & \sigma^{\mathrm{B}}~,(\rm pb) & R, & $\sigma^{\mathrm{UP}}$~(pb) \\
			\hline
			4.0076 & 482.0   & 0.682 & 1.044 & 12.50 & -,                             & -,                             & -,                              & $1.3  $  \\
			4.0855 & 52.86   & 0.766 & 1.051 & 11.20 & 1.0^{+0.5}_{-0.7}     ,\pm 0.1 & 5.2^{+2.6}_{-3.6}   ,\pm   0.3 & 1.04^{+0.66}_{-0.90}  ,\pm  0.07 & $20.2 $  \\
			4.1285 & 401.5   & 0.781 & 1.052 & 10.75 & 3.0^{+2.2}_{-1.5}     ,\pm 0.2 & 2.1^{+1.5}_{-1.0}   ,\pm   0.1 & 0.29^{+0.22}_{-0.15}  ,\pm  0.02 & $4.8  $  \\
			4.1574 & 408.7   & 0.787 & 1.053 & 10.86 & 5.8^{+3.1}_{-2.4}     ,\pm 0.4 & 3.9^{+2.1}_{-1.6}   ,\pm   0.2 & 0.41^{+0.22}_{-0.17}  ,\pm  0.03 & $-     $ \\
			4.1784 & 3194.5  & 0.789 & 1.054 & 11.18 & 54.5^{+7.9}_{-7.9}    ,\pm 3.5 & 4.5^{+0.7}_{-0.7}   ,\pm   0.3 & 0.45^{+0.07}_{-0.07}  ,\pm  0.03 & $-     $ \\
			4.1888 & 570.03  & 0.789 & 1.056 & 11.48 & 18.4^{+4.6}_{-4.6}    ,\pm 1.2 & 8.4^{+2.1}_{-2.1}   ,\pm   0.5 & 0.83^{+0.22}_{-0.22}  ,\pm  0.05 & $-     $ \\
			4.1989 & 526.0   & 0.788 & 1.056 & 11.58 & 11.4^{+3.7}_{-3.7}    ,\pm 0.7 & 5.6^{+1.8}_{-1.8}   ,\pm   0.4 & 0.40^{+0.13}_{-0.13}  ,\pm  0.03 & $-     $ \\
			4.2091 & 572.05  & 0.785 & 1.057 & 11.06 & 13.6^{+3.9}_{-3.9}    ,\pm 1.0 & 6.4^{+1.9}_{-1.9}   ,\pm   0.5 & 0.47^{+0.14}_{-0.14}  ,\pm  0.04 & $-     $ \\
			4.2186 & 569.2   & 0.778 & 1.056 & 11.69 & 24.1^{+5.5}_{-5.5}    ,\pm 1.8 & 10.9^{+2.5}_{-2.5}  ,\pm   0.8 & 0.57^{+0.14}_{-0.14}  ,\pm  0.04 & $-     $ \\
			4.2263 & 1100.9  & 0.766 & 1.056 & 11.82 & 42.3^{+7.0}_{-7.0}    ,\pm 3.2 & 9.9^{+1.6}_{-1.6}   ,\pm   0.8 & 0.50^{+0.09}_{-0.09}  ,\pm  0.04 & $-     $ \\
			4.2357 & 530.3   & 0.786 & 1.056 & 11.53 & 18.6^{+4.6}_{-4.6}    ,\pm 1.4 & 9.1^{+2.2}_{-2.2}   ,\pm   0.7 & 0.51^{+0.13}_{-0.13}  ,\pm  0.04 & $-     $ \\
			4.2436 & 593.98  & 0.859 & 1.056 & 11.17 & 26.1^{+5.4}_{-5.4}    ,\pm 2.0 & 10.8^{+2.2}_{-2.2}  ,\pm   0.8 & 0.65^{+0.14}_{-0.14}  ,\pm  0.05 & $-     $ \\
			4.2580 & 828.4   & 0.834 & 1.054 & 12.01 & 26.4^{+6.0}_{-6.0}    ,\pm 2.0 & 7.5^{+1.7}_{-1.7}   ,\pm   0.6 & 0.41^{+0.10}_{-0.10}  ,\pm  0.03 & $-     $ \\
			4.2668 & 531.1   & 0.820 & 1.053 & 11.83 & 14.5^{+4.6}_{-4.6}    ,\pm 1.1 & 6.6^{+2.1}_{-2.1}   ,\pm   0.5 & 0.31^{+0.10}_{-0.10}  ,\pm  0.02 & $-     $ \\
			4.2777 & 175.7   & 0.809 & 1.053 & 12.17 & 14.1^{+4.5}_{-4.5}    ,\pm 1.1 & 19.1^{+6.1}_{-6.1}  ,\pm   1.5 & 0.63^{+0.21}_{-0.21}  ,\pm  0.05 & $-     $ \\
			4.2879 & 502.4   & 0.803 & 1.053 & 11.02 & 23.1^{+6.0}_{-6.0}    ,\pm 1.8 & 12.2^{+3.1}_{-3.1}  ,\pm   0.9 & 0.43^{+0.11}_{-0.11}  ,\pm  0.03 & $-     $ \\
			4.3079 & 45.08   & 0.795 & 1.052 & 12.23 & 1.4^{+2.0}_{-1.0}     ,\pm 0.1 & 7.3^{+10.7}_{-5.5}  ,\pm   0.5 & 0.23^{+0.34}_{-0.18}  ,\pm  0.02 & $18.9 $  \\
			4.3121 & 501.2   & 0.794 & 1.052 & 11.97 & 41.1^{+6.9}_{-6.9}    ,\pm 2.9 & 20.3^{+3.4}_{-3.4}  ,\pm   1.5 & 0.53^{+0.09}_{-0.09}  ,\pm  0.04 & $-     $ \\
			4.3374 & 505.0   & 0.788 & 1.051 & 12.44 & 50.7^{+7.3}_{-7.3}    ,\pm 3.6 & 24.1^{+3.5}_{-3.5}  ,\pm   1.7 & 0.46^{+0.07}_{-0.07}  ,\pm  0.03 & $-     $ \\
			4.3583 & 543.9   & 0.791 & 1.051 & 13.15 & 62.9^{+8.5}_{-8.5}    ,\pm 4.5 & 26.2^{+3.5}_{-3.5}  ,\pm   1.9 & 0.46^{+0.06}_{-0.06}  ,\pm  0.03 & $-     $ \\
			4.3774 & 522.7   & 0.807 & 1.051 & 12.44 & 81.0^{+9.3}_{-9.3}    ,\pm 5.8 & 36.3^{+4.2}_{-4.2}  ,\pm   2.6 & 0.61^{+0.07}_{-0.07}  ,\pm  0.04 & $-     $ \\
			4.3874 & 55.57   & 0.826 & 1.051 & 12.78 & 6.6^{+3.1}_{-2.4}     ,\pm 0.5 & 26.4^{+12.4}_{-9.6} ,\pm   1.9 & 0.47^{+0.23}_{-0.18}  ,\pm  0.03 & $-     $ \\
			4.3965 & 507.8   & 0.848 & 1.051 & 12.22 & 59.1^{+8.2}_{-8.2}    ,\pm 3.8 & 26.4^{+3.7}_{-3.7}  ,\pm   1.7 & 0.47^{+0.07}_{-0.07}  ,\pm  0.03 & $-     $ \\
			4.4156 & 1090.7  & 0.913 & 1.052 & 12.28 & 103.7^{+10.5}_{-10.5} ,\pm 6.6 & 19.9^{+2.0}_{-2.0}  ,\pm   1.3 & 0.47^{+0.05}_{-0.05}  ,\pm  0.03 & $-     $ \\
			4.4362 & 569.9   & 1.008 & 1.054 & 11.22 & 42.8^{+7.0}_{-7.0}    ,\pm 2.7 & 15.6^{+2.5}_{-2.5}  ,\pm   1.0 & 0.47^{+0.08}_{-0.08}  ,\pm  0.03 & $-     $ \\
			4.4671 & 111.09  & 1.168 & 1.055 & 10.13 & 2.2^{+2.0}_{-1.2}     ,\pm 0.1 & 3.9^{+3.6}_{-2.2}   ,\pm   0.3 & 0.43^{+0.41}_{-0.27}  ,\pm  0.03 & $11.1 $  \\
			4.5271 & 112.12  & 1.367 & 1.054 & 8.37  & 2.0^{+1.0}_{-1.1}     ,\pm 0.1 & 3.6^{+1.9}_{-1.9}   ,\pm   0.2 & 0.46^{+0.26}_{-0.28}  ,\pm  0.03 & $9.8  $  \\
			4.5745 & 48.93   & 1.248 & 1.054 & 9.20  & -                              & -                              & -                               & $8.1  $ \\
			4.5995 & 586.9   & 1.123 & 1.055 & 10.26 & 13.1^{+3.9}_{-3.9}    ,\pm 1.0 & 4.5^{+1.4}_{-1.4}   ,\pm   0.4 & 0.35^{+0.11}_{-0.11}  ,\pm  0.03 & $-     $ \\
			4.6119 & 103.65  & 1.060 & 1.055 & 10.63 & 6.9^{+3.1}_{-2.2}     ,\pm 0.5 & 13.9^{+6.2}_{-4.5}  ,\pm   1.1 & 0.96^{+0.46}_{-0.36}  ,\pm  0.07 & $-     $ \\
			4.6280 & 521.53  & 0.986 & 1.054 & 11.28 & 25.3^{+5.8}_{-5.2}    ,\pm 2.0 & 10.2^{+2.4}_{-2.1}  ,\pm   0.8 & 0.51^{+0.12}_{-0.11}  ,\pm  0.04 & $-     $ \\
			4.6409 & 551.65  & 0.939 & 1.054 & 12.13 & 15.5^{+4.8}_{-4.1}    ,\pm 1.2 & 5.8^{+1.8}_{-1.5}   ,\pm   0.5 & 0.27^{+0.09}_{-0.07}  ,\pm  0.02 & $-     $ \\
			4.6612 & 529.43  & 0.889 & 1.054 & 12.98 & 24.0^{+5.5}_{-5.5}    ,\pm 1.9 & 9.2^{+2.1}_{-2.1}   ,\pm   0.7 & 0.38^{+0.09}_{-0.09}  ,\pm  0.03 & $-     $ \\
			4.6819 & 1667.39 & 0.877 & 1.054 & 13.33 & 100.3^{+10.8}_{-10.8} ,\pm 7.8 & 12.1^{+1.3}_{-1.3}  ,\pm   0.9 & 0.55^{+0.06}_{-0.06}  ,\pm  0.04 & $-     $ \\
			4.6988 & 535.54  & 0.898 & 1.055 & 13.71 & 31.7^{+6.1}_{-6.1}    ,\pm 2.5 & 11.3^{+2.2}_{-2.2}  ,\pm   0.9 & 0.60^{+0.12}_{-0.12}  ,\pm  0.05 & $-     $ \\
      4.7397 & 163.87  & 1.020 & 1.055 & 12.65 & 3.1^{+2.2}_{-1.6}     ,\pm 0.2 & 3.5^{+2.5}_{-1.7}   ,\pm   0.3 & -                               & $8.0  $  \\
			4.7501 & 366.55  & 1.057 & 1.055 & 12.48 & 16.0^{+4.2}_{-4.2}    ,\pm 1.2 & 7.8^{+2.0}_{-2.0}   ,\pm   0.6 & -                               & $-     $ \\
			4.7805 & 511.47  & 1.160 & 1.055 & 11.81 & 13.7^{+4.1}_{-4.1}    ,\pm 1.1 & 4.6^{+1.4}_{-1.4}   ,\pm   0.4 & -                               & $-     $ \\
			4.8431 & 525.16  & 1.317 & 1.056 & 9.64  & 10.9^{+3.5}_{-3.5}    ,\pm 0.8 & 3.9^{+1.2}_{-1.2}   ,\pm   0.3 & -                               & $-     $ \\
			4.9180 & 207.82  & 1.424 & 1.056 & 8.73  & 2.2^{+2.1}_{-1.4}     ,\pm 0.2 & 2.0^{+1.9}_{-1.2}   ,\pm   0.2 & -                               & $5.5  $  \\
			4.9509 & 159.28  & 1.451 & 1.056 & 8.45  & 1.0^{+0.5}_{-0.7}     ,\pm 0.1 & 1.2^{+0.6}_{-0.8}   ,\pm   0.1 & -                               & $4.2  $  \\
			\hline
			\hline
		\end{tabular}
	}
\end{table*}

\begin{table*}[htbp]
	\centering
	\caption{
		Results of the fits to the $\ee\to\pizpiz\psip$ dressed cross sections,
		where the first uncertainties are statistical and the second systematic.
	}
	\label{tab:sol_xsec_fit}

	\scalebox{0.85}
	{
		\begin{tabular}{lcccc}
			\hline
			\hline
			Parameter                      & ~~~~~~~Solution~I~~~~~~~               & ~~~~~~~Solution~II~~~~~~~         & ~~~~~~~Solution~III~~~~~~~        & ~~~~~~~Solution~IV~~~~~~~         \\
			\hline
			$M(Y4230$)~(MeV/$c^2$)          & \multicolumn{4}{c}{$4234.4 $~(fixed)}                                                                                                              \\
			$\Gamma^{\rm tot}(Y4230)$~(MeV) & \multicolumn{4}{c}{$17.6   $~(fixed)}                                                                                                              \\
      ${\mathcal B}\Gamma^{ee}(Y4230)$~ (eV)     & $ 0.81 \pm  0.07\pm 0.37 $                     & $ 0.02 \pm  0.01 \pm 0.01 $                & $ 0.87 \pm  0.08 \pm 0.43$                & $ 0.02 \pm  0.01 \pm 0.01$                \\
$M(Y4390$)~(MeV/$c^2$)          & \multicolumn{4}{c}{$4383.0  \pm 8.6 \pm 1.9$}                                                                                                              \\
			$\Gamma^{\rm tot}(Y4390)$~(MeV) & \multicolumn{4}{c}{$117.4   \pm 20.7 \pm 4.8$}                                                                                                             \\

			${\mathcal B}\Gamma^{ee}(Y4390)$~(eV)      & $ 3.18         \pm 1.21      \pm 0.57   $      & $ 2.92         \pm 1.07   \pm 0.36      $ & $ 9.90         \pm 1.42     \pm 0.76    $ & $ 9.03         \pm 1.22      \pm 0.09   $ \\
$M(Y4660$)~(MeV/$c^2$)          & \multicolumn{4}{c}{$4684.0  \pm 17.3 \pm 1.9$}\\
			$\Gamma^{\rm tot}(Y4660)$~(MeV) & \multicolumn{4}{c}{$119.5   \pm 47.1\pm9.1$}                                                                                                             \\
      ${\mathcal B}\Gamma^{ee}(Y4660)$~(eV)      & $ 0.80         \pm 0.73      \pm 0.14   $      & $ 0.79         \pm 0.70  \pm 0.13       $ & $ 5.36         \pm 1.60   \pm 0.39       $ & $ 5.31         \pm 1.58  \pm 0.23       $ \\
$\phi_{Y(4230)}~(\rm rad)$      & $ 2.03         \pm 0.25         $      & $ 6.14         \pm 0.41         $ & $ 1.23         \pm 0.18         $ & $ 5.35         \pm 0.52         $ \\
			$\phi_{Y(4660)}~(\rm rad)$      & $ 5.99         \pm 0.40         $      & $ 5.93         \pm 0.40         $ & $ 5.28         \pm 0.46         $ & $ 5.22         \pm 0.46         $ \\
			$\phi_{\rm cont}~(\rm rad)$     & $ 3.87         \pm 0.36         $      & $ 3.69         \pm 0.33         $ & $ 2.23         \pm 0.27         $ & $ 2.06         \pm 0.24         $ \\

$a~(\times10^{5})$ & $ 4.2 \pm 9.5 $ & $ 3.3 \pm 8.0 $ & $ 5.3 \pm 22.8 $ & $ 4.9 \pm 22.6 $ \\
			$n$                & $ 8.7 \pm 1.8 $ & $ 8.6 \pm 1.8 $ & $ 8.9 \pm 3.6  $ & $ 8.9 \pm 3.6  $ \\
			\hline
			\hline
		\end{tabular}
	}
\end{table*}

\section{Systematic Uncertainties}
\subsection{Systematic Uncertainties of the Cross Section Measurement}
Several sources of systematic uncertainties are considered in the
cross section measurement.
Due to the limited statistics of some data sets,
those with low statistics are estimated
with the nearest high-statistics data sets
($\sqrt{s}$ = 4.1784, 4.2263, 4.3583, 4.4156, 4.6819$~\gev$).

\begin{table}[htbp]
	\centering
	\caption{The summary of the relative systematic uncertainties~(in unit of~\%) at different energy points from different sources.
    The $L_{\text{int}}$ denotes integral luminosity, ${\mathcal B}$ denotes branching fraction,  Track I and Track II
		denote charged and neutral tracks, respectively. The sources with ``*" are the common systematic
		uncertainties for all the energy points.}
	\label{tab:sys_err}
	\scalebox{0.92}{
		\begin{tabular}{l|c|c|c|c|c}
			\hline
			\hline
$\sqrt{s}$~(GeV) & ~4.1784~ & ~4.2263~ & ~4.3583~ & ~4.4156~ & ~4.6819~ \\
			\hline
      $L_{\text{int}}^{*}$ & 1.0 & 1.0 & 1.0 & 1.0 & 1.0 \\
      ${\mathcal B}^{*}$              & 1.0 & 1.0 & 1.0 & 1.0 & 1.0 \\
			Track I$^{*}$        & 4.0 & 4.0 & 4.0 & 4.0 & 4.0 \\
			Track II$^{*}$       & 4.0 & 4.0 & 4.0 & 4.0 & 4.0 \\
			ISR correction       & 0.9 & 2.3 & 1.2 & 1.8 & 3.7 \\
			$J/\psi$ mass        & 0.6 & 0.6 & 0.2 & 0.1 & 0.8 \\
			$\pi^0$ mass         & 0.4 & 0.6 & 0.2 & 0.1 & 0.1 \\
			Kinematic fit        & 0.5 & 0.5 & 0.4 & 0.5 & 0.4 \\
			MC model             & 1.2 & 1.9 & 1.0 & 1.3 & 1.7 \\
			Signal shape         & 0.6 & 0.5 & 0.5 & 0.2 & 0.6 \\
			Background shape~     & 0.0 & 0.1 & 0.0 & 0.2 & 0.2 \\
			Others$^{*}$         & 1.0 & 1.0 & 1.0 & 1.0 & 1.0 \\
			Total                & 6.2 & 6.7 & 6.2 & 6.3 & 7.2 \\

			\hline
			\hline
		\end{tabular}
	}
\end{table}

The uncertainty of the integrated luminosity is
1.0\%~\cite{CPC46113003, CPC39093001, CPC46113002}.
The uncertainties of the branching fractions from the intermediate
decays $\piz\to\gam\gam~(0.034\%)$, $\psip\to\pipi\jpsi~(0.34\%)$ and $\jpsi\to\llep~(0.045\%)$
are taken from the PDG~\cite{PRD110030001}.
The uncertainties of the charged track reconstruction and particle identification are both
 1.0\% for each charged track~\cite{PRL110252001,PRD83112005},
and 4.0\% is assigned.
The uncertainty of the photon detection is 1.0\% per photon~\cite{PRD81052005},
and 4.0\% is assigned for two $\pi^0$ mesons.
The uncertainty of the radiative correction includes two parts.
The first one stems from the precision of the ISR
calculation in the generator {\sc kkmc}~\cite{CPC38083001}, which is 0.5\%.
The other one is from $(1+\delta^{\mathrm{ISR}})$ and $\epsilon$
in Eq.~\ref{eq:xsec}, relying on the
input line shape of cross sections. To estimate the uncertainty, the
$(1+\delta^{\mathrm{ISR}})\cdot\epsilon$ is evaluated for 100 times by varying
the input line shape within the uncertainties and the covariance matrix
obtained from the nominal cross section line shape fit. The standard deviation
of the $(1+\delta^{\mathrm{ISR}})\cdot\epsilon$ distribution is taken as the
systematic uncertainty of the radiative correction.
The uncertainty associated with the $\jpsi$ mass requirement is
estimated by smearing the $M(\llep)$ distribution of the MC sample
with a Gaussian function, which represents the resolution difference
between data and MC simulation, and the resulting signal efficiency variation
is taken as the systematic uncertainty.
The same procedure is applied to the $\pi^0$ mass requirement
to estimate the related systematic uncertainty.
For the systematic uncertainty from the kinematic fit, the
helix parameters of the charged tracks are corrected to
make the pull distribution of MC sample more consistent with
the data and the signal efficiency is re-evaluated, and the
resultant difference is taken as the uncertainty.
To estimate the systematic uncertainty of the MC modeling,
the signal efficiency is re-evaluated using 200 regenerated MC samples
in which the PWA parameters are varied within their uncertainties,
while taking into account the correlations among these parameters through their full covariance matrix.
The standard deviation of the signal efficiency distribution
is taken as the systematic uncertainty.
The systematic uncertainties related to the fit procedure
are estimated by replacing the first-order polynomial function
by a second-order polynomial
function for the background shape, and varying the
resolution of the convolved Gaussian function for the signal
shape by one standard deviation.
Following the ``Barlow-test" method in Refs.~\cite{arXivhepex0207026,N6066469},
the systematic uncertainty related to the fit range is
found to be negligible.
The uncertainties from the
other selections, trigger simulation, event start time determination,
and FSR simulation and other
sources, are conservatively taken as 1.0\%.
All the systematic uncertainties are summarized in Table~\ref{tab:sys_err}.
Assuming all sources of systematic uncertainties are
independent, the total uncertainties in the cross section
measurement are obtained by summing the individual uncertainties
in quadrature, which range from 6.2\% to 7.2\%.

\begin{table*}[htbp]
	\centering
	\caption{The summary of absolute systematic uncertainties of
		resonant parameters in the fit to the cross sections of the
		$\ee\to\pizpiz\psip$ process, including the c.m. energy $\sqrt{s}$,
		the energy spread~(Ene. spread),
		the fit model, the parameters of $Y(4230)$
		and the systematic uncertainty in the cross section measurement (Cross section).
		The ``..." represents that the uncertainty is negligible.}
	\label{tab:sys_unc_res}
	\scalebox{1.0}
	{
		\begin{tabular}{l|ccccccc}
			\hline
			\hline
			Source                                                            & ~~Solution~~ & ~~$\sqrt{s}$~~ & ~~Ene. spread~~ & ~~Model~~ & ~~$\YI$~~ & ~~Cross section~~ & ~~Total~~ \\
			\hline
			$M(Y(4390))$                         (MeV/$c^2$)                  & -            & 0.8            & 0.0             & 0.4       & 1.5       & 0.4               & 1.9       \\
			$\Gamma(Y(4390))$                    (MeV)                        & -            & ...            & 0.0             & 0.8       & 4.7       & 0.8               & 4.8       \\
			$M(Y(4660))$                         (MeV/$c^2$)                  & -            & 0.8            & 0.0             & 0.1       & 0.4       & 1.7               & 1.9       \\
			$\Gamma(Y(4660))$                    (MeV)                        & -            & ...            & 0.0             & 3.0       & 5.5       & 6.6               & 9.1       \\
			\hline
      \multirow{4}{*}{${\mathcal B}\Gamma^{e^{+}e^{-}}(Y(4230))$ (eV)}~ & I            & ...            & 0.00            & 0.00      & 0.37      & 0.00              & 0.37      \\
                                                                        & II           & ...            & 0.00            & 0.00      & 0.01      & 0.00              & 0.01      \\
                                                                        & III          & ...            & 0.00            & 0.00      & 0.43      & 0.00              & 0.43      \\
                                                                        & IV           & ...            & 0.00            & 0.00      & 0.01      & 0.00              & 0.01      \\
			\hline
      \multirow{4}{*}{${\mathcal B}\Gamma^{e^{+}e^{-}}(Y(4390))$ (eV)}~ & I            & ...            & 0.00            & 0.10      & 0.55      & 0.10              & 0.57      \\
                                                                        & II           & ...            & 0.00            & 0.08      & 0.34      & 0.09              & 0.36      \\
                                                                        & III          & ...            & 0.01            & 0.06      & 0.76      & 0.02              & 0.76      \\
                                                                        & IV           & ...            & 0.01            & 0.02      & 0.08      & 0.01              & 0.09      \\
			\hline
      \multirow{4}{*}{${\mathcal B}\Gamma^{e^{+}e^{-}}(Y(4660))$ (eV)}~ & I            & ...            & 0.00            & 0.05      & 0.09      & 0.09              & 0.14      \\
                                                                        & II           & ...            & 0.00            & 0.05      & 0.09      & 0.08              & 0.13      \\
                                                                        & III          & ...            & 0.01            & 0.14      & 0.33      & 0.15              & 0.39      \\
                                                                        & IV           & ...            & 0.01            & 0.12      & 0.11      & 0.15              & 0.23      \\
			\hline
			\hline
		\end{tabular}
	}
\end{table*}

\subsection{Systematic Uncertainties of Resonant Parameter Measurement}
The systematic uncertainties of the resonant parameters from the fit to the
cross section line shape are as follows.
The systematic uncertainty associated with the collision energy is
conservatively estimated to be 0.8~$\mev$~\cite{CPC40063001,CPC45103001}.
It is common for all data samples and causes a global shift for the
masses of the resonances.
The uncertainty due to the energy spread is estimated by convolving the
fit probability density function with a Gaussian function with a resolution of 1.6~$\mev$, which is
the energy spread determined by the beam energy measurement system~\cite{NIMA6592129}.
The uncertainty from the statistical uncertainties of the cross sections is
estimated by incorporating the correlated and uncorrelated systematic uncertainties of
the measured cross sections in the fit following the likelihood construction method
in Ref.~\cite{PRD109092012}.
In the nominal fit, the mass and width of $\YI$ are fixed to the
values from Ref.~\cite{PRD104052012}. To estimate the relevant systematic uncertainty, the
mass and width of $\YI$ are varied within their uncertainties. The resultant differences of
each parameter
are taken as the systematic uncertainties.
To estimate the uncertainty related to the parameterization of the non-resonant part in
the fit model, the non-resonant part in Eq.~\ref{eq:fitmodel}
is replaced by a function of $\psi_{\mathrm{cont}}=\sqrt{\Phi(\sqrt{s}) e^{-p_0 u} p_1}$
in Ref.~\cite{PRD86051102,PRL118092001}
with $p_0$ and $p_1$ as free parameters, and $u=\sqrt{s}-(2M(\pi^{\pm})+M(\psip))$.
The cross section line shape fit is performed with the above scenario individually and the resulting differences are taken as the systematic uncertainties
as shown in Table~\ref{tab:sys_unc_res}.
The total systematic
uncertainty is obtained by quadratically summing the individual systematic uncertainties
in quadrature, under the assumption that
they are uncorrelated.

\section{SUMMARY}

The cross sections of the process $\ee\to\piz\piz\psip$ are measured at c.m.~energies between
4.008 and 4.951~$\gev$ using data samples with a total
integrated luminosity of $\lumtot$ collected with the BESIII detector
operating at the BEPCII. The measured cross sections are found to be
approximately one-half of those of the charged channel
$\ee\to\pipi\psip$~\cite{PRD104052012}, which are consistent with the
expectation from isospin symmetry.

The dressed cross sections are fitted with a coherent sum of three resonant
structures, $\YI$, $\YII$~and $\YIII$, as well as the same non-resonant component as used in the charged channel.
The statistical significances of $\YII$ and $\YIII$ are larger than $5\sigma$, which confirm
the existence of these two states in the $\ee\to\piz\piz\psip$ process.
The mass and width of $\YII$ are measured to be $\massTwo$ MeV/$c^2$,
$\widthTwo$ MeV, and those of $\YIII$ are
$\massThree$ MeV/$c^2$, $\widthThree$ MeV, where
the first uncertainties are statistical and the second systematic.
The parameters of the resonant structures are consistent with the
results from the charged channel within uncertainties.
The results from this work provide valuable information for the studies of
the charmonium-like states and the nature of the $XYZ$ particles.

\section*{ACKNOWLEDGEMENTS}
The BESIII Collaboration thanks the staff of BEPCII (https://cstr.cn/31109.02.BEPC), the IHEP computing center
and the supercomputing center of the University of Science and Technology of China (USTC)
for their strong support. 
This work is supported in part by
National Key R\&D Program of China under Contracts Nos. 2025YFA1613900, 2023YFA1609400, 2023YFA1606000, 2023YFA1606704;
National Natural Science Foundation of China (NSFC) under Contracts Nos.
12122509, 12105276,
11635010, 11935015, 11935016, 11935018, 12025502, 12035009, 12035013, 12061131003, 12192260, 12192261,
12192262, 12192263, 12192264, 12192265, 12221005, 12225509, 12235017, 12361141819;
the Chinese Academy of Sciences (CAS) Large-Scale Scientific Facility Program;
the Strategic Priority Research Program of Chinese Academy of Sciences under Contract No. XDA0480600;
Joint Large-Scale Scientific Facility Funds of the NSFC and CAS under Contracts Nos.
U2032111;
CAS under Contract No. YSBR-101; 100 Talents Program of CAS;
The Institute of Nuclear and Particle Physics (INPAC) and Shanghai Key Laboratory for Particle Physics and Cosmology;
ERC under Contract No. 758462; German Research Foundation DFG under Contract No. FOR5327;
Istituto Nazionale di Fisica Nucleare, Italy; Knut and Alice Wallenberg Foundation under Contracts Nos. 2021.0174, 2021.0299;
Ministry of Development of Turkey under Contract No. DPT2006K-120470; National Research Foundation of Korea under Contract No. NRF-2022R1A2C1092335;
National Science and Technology fund of Mongolia; Polish National Science Centre under Contract No. 2024/53/B/ST2/00975;
STFC (United Kingdom); Swedish Research Council under Contract No. 2019.04595; U. S. Department of Energy under Contract
No. DE-FG02-05ER41374


\begin{thebibliography}{60}%
\makeatletter
\providecommand \@ifxundefined [1]{%
 \@ifx{#1\undefined}
}%
\providecommand \@ifnum [1]{%
 \ifnum #1\expandafter \@firstoftwo
 \else \expandafter \@secondoftwo
 \fi
}%
\providecommand \@ifx [1]{%
 \ifx #1\expandafter \@firstoftwo
 \else \expandafter \@secondoftwo
 \fi
}%
\providecommand \natexlab [1]{#1}%
\providecommand \enquote  [1]{``#1''}%
\providecommand \bibnamefont  [1]{#1}%
\providecommand \bibfnamefont [1]{#1}%
\providecommand \citenamefont [1]{#1}%
\providecommand \href@noop [0]{\@secondoftwo}%
\providecommand \href [0]{\begingroup \@sanitize@url \@href}%
\providecommand \@href[1]{\@@startlink{#1}\@@href}%
\providecommand \@@href[1]{\endgroup#1\@@endlink}%
\providecommand \@sanitize@url [0]{\catcode `\\12\catcode `\$12\catcode `\&12\catcode `\#12\catcode `\^12\catcode `\_12\catcode `\%12\relax}%
\providecommand \@@startlink[1]{}%
\providecommand \@@endlink[0]{}%
\providecommand \url  [0]{\begingroup\@sanitize@url \@url }%
\providecommand \@url [1]{\endgroup\@href {#1}{\urlprefix }}%
\providecommand \urlprefix  [0]{URL }%
\providecommand \Eprint [0]{\href }%
\providecommand \doibase [0]{http://dx.doi.org/}%
\providecommand \selectlanguage [0]{\@gobble}%
\providecommand \bibinfo  [0]{\@secondoftwo}%
\providecommand \bibfield  [0]{\@secondoftwo}%
\providecommand \translation [1]{[#1]}%
\providecommand \BibitemOpen [0]{}%
\providecommand \bibitemStop [0]{}%
\providecommand \bibitemNoStop [0]{.\EOS\space}%
\providecommand \EOS [0]{\spacefactor3000\relax}%
\providecommand \BibitemShut  [1]{\csname bibitem#1\endcsname}%
\let\auto@bib@innerbib\@empty
\bibitem [{\citenamefont {Barnes}\ \emph {et~al.}(2005)\citenamefont {Barnes}, \citenamefont {Godfrey},\ and\ \citenamefont {Swanson}}]{PRD72054026}%
  \BibitemOpen
  \bibfield  {author} {\bibinfo {author} {\bibfnamefont {T.}~\bibnamefont {Barnes}}, \bibinfo {author} {\bibfnamefont {S.}~\bibnamefont {Godfrey}}, \ and\ \bibinfo {author} {\bibfnamefont {E.~S.}\ \bibnamefont {Swanson}},\ }\href {\doibase 10.1103/PhysRevD.72.054026} {\bibfield  {journal} {\bibinfo  {journal} {Physical Review D}\ }\textbf {\bibinfo {volume} {72}},\ \bibinfo {pages} {054026} (\bibinfo {year} {2005})}\BibitemShut {NoStop}%
\bibitem [{\citenamefont {Chen}\ \emph {et~al.}(2016)\citenamefont {Chen}, \citenamefont {Chen}, \citenamefont {Liu},\ and\ \citenamefont {Zhu}}]{PR6391121}%
  \BibitemOpen
  \bibfield  {author} {\bibinfo {author} {\bibfnamefont {H.}~\bibnamefont {Chen}}, \bibinfo {author} {\bibfnamefont {W.}~\bibnamefont {Chen}}, \bibinfo {author} {\bibfnamefont {X.}~\bibnamefont {Liu}}, \ and\ \bibinfo {author} {\bibfnamefont {S.}~\bibnamefont {Zhu}},\ }\href {\doibase 10.1016/j.physrep.2016.05.004} {\bibfield  {journal} {\bibinfo  {journal} {Phys. Rept.}\ }\textbf {\bibinfo {volume} {639}},\ \bibinfo {pages} {1} (\bibinfo {year} {2016})}\BibitemShut {NoStop}%
\bibitem [{\citenamefont {Guo}\ \emph {et~al.}(2018)\citenamefont {Guo}, \citenamefont {Hanhart}, \citenamefont {Mei\ss{}ner}, \citenamefont {Wang}, \citenamefont {Zhao},\ and\ \citenamefont {Zou}}]{RMP90015004}%
  \BibitemOpen
  \bibfield  {author} {\bibinfo {author} {\bibfnamefont {F.~K.}\ \bibnamefont {Guo}}, \bibinfo {author} {\bibfnamefont {C.}~\bibnamefont {Hanhart}}, \bibinfo {author} {\bibfnamefont {U.~G.}\ \bibnamefont {Mei\ss{}ner}}, \bibinfo {author} {\bibfnamefont {Q.}~\bibnamefont {Wang}}, \bibinfo {author} {\bibfnamefont {Q.}~\bibnamefont {Zhao}}, \ and\ \bibinfo {author} {\bibfnamefont {B.~S.}\ \bibnamefont {Zou}},\ }\href {\doibase 10.1103/RevModPhys.90.015004} {\bibfield  {journal} {\bibinfo  {journal} {Rev. Mod. Phys.}\ }\textbf {\bibinfo {volume} {90}},\ \bibinfo {pages} {015004} (\bibinfo {year} {2018})}\BibitemShut {NoStop}%
\bibitem [{\citenamefont {Brambilla}\ \emph {et~al.}(2020)\citenamefont {Brambilla} \emph {et~al.}}]{PR8731154}%
  \BibitemOpen
  \bibfield  {author} {\bibinfo {author} {\bibfnamefont {N.}~\bibnamefont {Brambilla}} \emph {et~al.},\ }\href {\doibase 10.1016/j.physrep.2020.05.001} {\bibfield  {journal} {\bibinfo  {journal} {Phys. Rept.}\ }\textbf {\bibinfo {volume} {873}},\ \bibinfo {pages} {1} (\bibinfo {year} {2020})}\BibitemShut {NoStop}%
\bibitem [{\citenamefont {Wang}(2026)}]{FPB21016300}%
  \BibitemOpen
  \bibfield  {author} {\bibinfo {author} {\bibfnamefont {Z.~G.}\ \bibnamefont {Wang}},\ }\href {\doibase 10.15302/frontphys.2026.016300} {\bibfield  {journal} {\bibinfo  {journal} {Front. Phys. (Beijing)}\ }\textbf {\bibinfo {volume} {21}},\ \bibinfo {pages} {016300} (\bibinfo {year} {2026})}\BibitemShut {NoStop}%
\bibitem [{\citenamefont {Wang}\ \emph {et~al.}(2025)\citenamefont {Wang}, \citenamefont {Liu},\ and\ \citenamefont {Gao}}]{arXiv250215117}%
  \BibitemOpen
  \bibfield  {author} {\bibinfo {author} {\bibfnamefont {X.}~\bibnamefont {Wang}}, \bibinfo {author} {\bibfnamefont {X.}~\bibnamefont {Liu}}, \ and\ \bibinfo {author} {\bibfnamefont {Y.}~\bibnamefont {Gao}},\ }\href {\doibase 10.48550/arXiv.2502.15117} {\enquote {\bibinfo {title} {Colloquium: {Hadron} {Production} in {Open}-charm {Meson} {Pair} at $e^{+}e^{-}$ {Collider}},}\ } (\bibinfo {year} {2025}),\ \bibinfo {note} {arXiv:2502.15117 [hep-ex]}\BibitemShut {NoStop}%
\bibitem [{\citenamefont {Nakamura}\ \emph {et~al.}(2025)\citenamefont {Nakamura}, \citenamefont {Li}, \citenamefont {Peng}, \citenamefont {Sun},\ and\ \citenamefont {Zhou}}]{PRD112054027}%
  \BibitemOpen
  \bibfield  {author} {\bibinfo {author} {\bibfnamefont {S.}~\bibnamefont {Nakamura}}, \bibinfo {author} {\bibfnamefont {X.~H.}\ \bibnamefont {Li}}, \bibinfo {author} {\bibfnamefont {H.~P.}\ \bibnamefont {Peng}}, \bibinfo {author} {\bibfnamefont {Z.~T.}\ \bibnamefont {Sun}}, \ and\ \bibinfo {author} {\bibfnamefont {X.~R.}\ \bibnamefont {Zhou}},\ }\href {\doibase 10.1103/fch8-xwb8} {\bibfield  {journal} {\bibinfo  {journal} {Phys. Rev. D}\ }\textbf {\bibinfo {volume} {112}},\ \bibinfo {pages} {054027} (\bibinfo {year} {2025})}\BibitemShut {NoStop}%
\bibitem [{\citenamefont {Aubert}\ \emph {et~al.}(2005)\citenamefont {Aubert} \emph {et~al.}}]{PRL95142001}%
  \BibitemOpen
  \bibfield  {author} {\bibinfo {author} {\bibfnamefont {B.}~\bibnamefont {Aubert}} \emph {et~al.} (\bibinfo {collaboration} {BABAR Collaboration}),\ }\href {\doibase 10.1103/PhysRevLett.95.142001} {\bibfield  {journal} {\bibinfo  {journal} {Phys. Rev. Lett.}\ }\textbf {\bibinfo {volume} {95}},\ \bibinfo {pages} {142001} (\bibinfo {year} {2005})}\BibitemShut {NoStop}%
\bibitem [{\citenamefont {Yuan}\ \emph {et~al.}(2007)\citenamefont {Yuan} \emph {et~al.}}]{PRL99182004}%
  \BibitemOpen
  \bibfield  {author} {\bibinfo {author} {\bibfnamefont {C.~Z.}\ \bibnamefont {Yuan}} \emph {et~al.} (\bibinfo {collaboration} {Belle Collaboration}),\ }\href {\doibase 10.1103/PhysRevLett.99.182004} {\bibfield  {journal} {\bibinfo  {journal} {Phys. Rev. Lett.}\ }\textbf {\bibinfo {volume} {99}},\ \bibinfo {pages} {182004} (\bibinfo {year} {2007})}\BibitemShut {NoStop}%
\bibitem [{\citenamefont {Lees}\ \emph {et~al.}(2012)\citenamefont {Lees} \emph {et~al.}}]{PRD86051102}%
  \BibitemOpen
  \bibfield  {author} {\bibinfo {author} {\bibfnamefont {J.~P.}\ \bibnamefont {Lees}} \emph {et~al.} (\bibinfo {collaboration} {BABAR Collaboration}),\ }\href {\doibase 10.1103/PhysRevD.86.051102} {\bibfield  {journal} {\bibinfo  {journal} {Phys. Rev. D}\ }\textbf {\bibinfo {volume} {86}},\ \bibinfo {pages} {051102} (\bibinfo {year} {2012})}\BibitemShut {NoStop}%
\bibitem [{\citenamefont {Liu}\ \emph {et~al.}(2013)\citenamefont {Liu} \emph {et~al.}}]{PRL110252002}%
  \BibitemOpen
  \bibfield  {author} {\bibinfo {author} {\bibfnamefont {Z.~Q.}\ \bibnamefont {Liu}} \emph {et~al.} (\bibinfo {collaboration} {Belle Collaboration}),\ }\href {\doibase 10.1103/PhysRevLett.110.252002} {\bibfield  {journal} {\bibinfo  {journal} {Phys. Rev. Lett.}\ }\textbf {\bibinfo {volume} {110}},\ \bibinfo {pages} {252002} (\bibinfo {year} {2013})},\ \bibinfo {note} {[Erratum: Phys.Rev.Lett. 111, 019901 (2013)]}\BibitemShut {NoStop}%
\bibitem [{\citenamefont {Aubert}\ \emph {et~al.}(2007)\citenamefont {Aubert} \emph {et~al.}}]{PRL98212001}%
  \BibitemOpen
  \bibfield  {author} {\bibinfo {author} {\bibfnamefont {B.}~\bibnamefont {Aubert}} \emph {et~al.} (\bibinfo {collaboration} {BABAR Collaboration}),\ }\href {\doibase 10.1103/PhysRevLett.98.212001} {\bibfield  {journal} {\bibinfo  {journal} {Phys. Rev. Lett.}\ }\textbf {\bibinfo {volume} {98}},\ \bibinfo {pages} {212001} (\bibinfo {year} {2007})}\BibitemShut {NoStop}%
\bibitem [{\citenamefont {Lees}\ \emph {et~al.}(2014)\citenamefont {Lees} \emph {et~al.}}]{PRD89111103}%
  \BibitemOpen
  \bibfield  {author} {\bibinfo {author} {\bibfnamefont {J.~P.}\ \bibnamefont {Lees}} \emph {et~al.} (\bibinfo {collaboration} {BABAR Collaboration}),\ }\href {\doibase 10.1103/PhysRevD.89.111103} {\bibfield  {journal} {\bibinfo  {journal} {Phys. Rev. D}\ }\textbf {\bibinfo {volume} {89}},\ \bibinfo {pages} {111103} (\bibinfo {year} {2014})}\BibitemShut {NoStop}%
\bibitem [{\citenamefont {Wang}\ \emph {et~al.}(2007)\citenamefont {Wang} \emph {et~al.}}]{PRL99142002}%
  \BibitemOpen
  \bibfield  {author} {\bibinfo {author} {\bibfnamefont {X.~L.}\ \bibnamefont {Wang}} \emph {et~al.} (\bibinfo {collaboration} {Belle Collaboration}),\ }\href {\doibase 10.1103/PhysRevLett.99.142002} {\bibfield  {journal} {\bibinfo  {journal} {Phys. Rev. Lett.}\ }\textbf {\bibinfo {volume} {99}},\ \bibinfo {pages} {142002} (\bibinfo {year} {2007})}\BibitemShut {NoStop}%
\bibitem [{\citenamefont {Wang}\ \emph {et~al.}(2015)\citenamefont {Wang} \emph {et~al.}}]{PRD91112007}%
  \BibitemOpen
  \bibfield  {author} {\bibinfo {author} {\bibfnamefont {X.~L.}\ \bibnamefont {Wang}} \emph {et~al.} (\bibinfo {collaboration} {Belle Collaboration}),\ }\href {\doibase 10.1103/PhysRevD.91.112007} {\bibfield  {journal} {\bibinfo  {journal} {Phys. Rev. D}\ }\textbf {\bibinfo {volume} {91}},\ \bibinfo {pages} {112007} (\bibinfo {year} {2015})}\BibitemShut {NoStop}%
\bibitem [{\citenamefont {Ablikim}\ \emph {et~al.}(2017{\natexlab{a}})\citenamefont {Ablikim} \emph {et~al.}}]{PRL118092001}%
  \BibitemOpen
  \bibfield  {author} {\bibinfo {author} {\bibfnamefont {M.}~\bibnamefont {Ablikim}} \emph {et~al.} (\bibinfo {collaboration} {BESIII Collaboration}),\ }\href {\doibase 10.1103/PhysRevLett.118.092001} {\bibfield  {journal} {\bibinfo  {journal} {Phys. Rev. Lett.}\ }\textbf {\bibinfo {volume} {118}},\ \bibinfo {pages} {092001} (\bibinfo {year} {2017}{\natexlab{a}})}\BibitemShut {NoStop}%
\bibitem [{\citenamefont {Ablikim}\ \emph {et~al.}(2020{\natexlab{a}})\citenamefont {Ablikim} \emph {et~al.}}]{PRD102012009}%
  \BibitemOpen
  \bibfield  {author} {\bibinfo {author} {\bibfnamefont {M.}~\bibnamefont {Ablikim}} \emph {et~al.} (\bibinfo {collaboration} {BESIII Collaboration}),\ }\href {\doibase 10.1103/PhysRevD.102.012009} {\bibfield  {journal} {\bibinfo  {journal} {Phys. Rev. D}\ }\textbf {\bibinfo {volume} {102}},\ \bibinfo {pages} {012009} (\bibinfo {year} {2020}{\natexlab{a}})}\BibitemShut {NoStop}%
\bibitem [{\citenamefont {Ablikim}\ \emph {et~al.}(2023)\citenamefont {Ablikim} \emph {et~al.}}]{PRD107092005}%
  \BibitemOpen
  \bibfield  {author} {\bibinfo {author} {\bibfnamefont {M.}~\bibnamefont {Ablikim}} \emph {et~al.} (\bibinfo {collaboration} {BESIII Collaboration}),\ }\href {\doibase 10.1103/PhysRevD.107.092005} {\bibfield  {journal} {\bibinfo  {journal} {Phys. Rev. D}\ }\textbf {\bibinfo {volume} {107}},\ \bibinfo {pages} {092005} (\bibinfo {year} {2023})}\BibitemShut {NoStop}%
\bibitem [{\citenamefont {Ablikim}\ \emph {et~al.}(2022{\natexlab{a}})\citenamefont {Ablikim} \emph {et~al.}}]{CPC46111002}%
  \BibitemOpen
  \bibfield  {author} {\bibinfo {author} {\bibfnamefont {M.}~\bibnamefont {Ablikim}} \emph {et~al.} (\bibinfo {collaboration} {BESIII Collaboration}),\ }\href {\doibase 10.1088/1674-1137/ac945c} {\bibfield  {journal} {\bibinfo  {journal} {Chin. Phys. C}\ }\textbf {\bibinfo {volume} {46}},\ \bibinfo {pages} {111002} (\bibinfo {year} {2022}{\natexlab{a}})}\BibitemShut {NoStop}%
\bibitem [{\citenamefont {Ablikim}\ \emph {et~al.}(2017{\natexlab{b}})\citenamefont {Ablikim} \emph {et~al.}}]{PRD96032004}%
  \BibitemOpen
  \bibfield  {author} {\bibinfo {author} {\bibfnamefont {M.}~\bibnamefont {Ablikim}} \emph {et~al.} (\bibinfo {collaboration} {BESIII Collaboration}),\ }\href {\doibase 10.1103/PhysRevD.96.032004} {\bibfield  {journal} {\bibinfo  {journal} {Phys. Rev. D}\ }\textbf {\bibinfo {volume} {96}},\ \bibinfo {pages} {032004} (\bibinfo {year} {2017}{\natexlab{b}})},\ \bibinfo {note} {[Erratum: Phys.Rev.D 99, 019903 (2019)]}\BibitemShut {NoStop}%
\bibitem [{\citenamefont {Ablikim}\ \emph {et~al.}(2021{\natexlab{a}})\citenamefont {Ablikim} \emph {et~al.}}]{PRD104052012}%
  \BibitemOpen
  \bibfield  {author} {\bibinfo {author} {\bibfnamefont {M.}~\bibnamefont {Ablikim}} \emph {et~al.} (\bibinfo {collaboration} {BESIII Collaboration}),\ }\href {\doibase 10.1103/PhysRevD.104.052012} {\bibfield  {journal} {\bibinfo  {journal} {Phys. Rev. D}\ }\textbf {\bibinfo {volume} {104}},\ \bibinfo {pages} {052012} (\bibinfo {year} {2021}{\natexlab{a}})}\BibitemShut {NoStop}%
\bibitem [{\citenamefont {Ablikim}\ \emph {et~al.}(2017{\natexlab{c}})\citenamefont {Ablikim} \emph {et~al.}}]{PRL118092002}%
  \BibitemOpen
  \bibfield  {author} {\bibinfo {author} {\bibfnamefont {M.}~\bibnamefont {Ablikim}} \emph {et~al.} (\bibinfo {collaboration} {BESIII Collaboration}),\ }\href {\doibase 10.1103/PhysRevLett.118.092002} {\bibfield  {journal} {\bibinfo  {journal} {Phys. Rev. Lett.}\ }\textbf {\bibinfo {volume} {118}},\ \bibinfo {pages} {092002} (\bibinfo {year} {2017}{\natexlab{c}})}\BibitemShut {NoStop}%
\bibitem [{\citenamefont {Ablikim}\ \emph {et~al.}(2015{\natexlab{a}})\citenamefont {Ablikim} \emph {et~al.}}]{PRL114092003}%
  \BibitemOpen
  \bibfield  {author} {\bibinfo {author} {\bibfnamefont {M.}~\bibnamefont {Ablikim}} \emph {et~al.} (\bibinfo {collaboration} {BESIII Collaboration}),\ }\href {\doibase 10.1103/PhysRevLett.114.092003} {\bibfield  {journal} {\bibinfo  {journal} {Phys. Rev. Lett.}\ }\textbf {\bibinfo {volume} {114}},\ \bibinfo {pages} {092003} (\bibinfo {year} {2015}{\natexlab{a}})}\BibitemShut {NoStop}%
\bibitem [{\citenamefont {Ablikim}\ \emph {et~al.}(2019{\natexlab{a}})\citenamefont {Ablikim} \emph {et~al.}}]{PRD99091103}%
  \BibitemOpen
  \bibfield  {author} {\bibinfo {author} {\bibfnamefont {M.}~\bibnamefont {Ablikim}} \emph {et~al.} (\bibinfo {collaboration} {BESIII Collaboration}),\ }\href {\doibase 10.1103/PhysRevD.99.091103} {\bibfield  {journal} {\bibinfo  {journal} {Phys. Rev. D}\ }\textbf {\bibinfo {volume} {99}},\ \bibinfo {pages} {091103} (\bibinfo {year} {2019}{\natexlab{a}})}\BibitemShut {NoStop}%
\bibitem [{\citenamefont {Ablikim}\ \emph {et~al.}(2019{\natexlab{b}})\citenamefont {Ablikim} \emph {et~al.}}]{PRL122102002}%
  \BibitemOpen
  \bibfield  {author} {\bibinfo {author} {\bibfnamefont {M.}~\bibnamefont {Ablikim}} \emph {et~al.} (\bibinfo {collaboration} {BESIII Collaboration}),\ }\href {\doibase 10.1103/PhysRevLett.122.102002} {\bibfield  {journal} {\bibinfo  {journal} {Phys. Rev. Lett.}\ }\textbf {\bibinfo {volume} {122}},\ \bibinfo {pages} {102002} (\bibinfo {year} {2019}{\natexlab{b}})}\BibitemShut {NoStop}%
\bibitem [{\citenamefont {Ablikim}\ \emph {et~al.}(2020{\natexlab{b}})\citenamefont {Ablikim} \emph {et~al.}}]{PRD102031101}%
  \BibitemOpen
  \bibfield  {author} {\bibinfo {author} {\bibfnamefont {M.}~\bibnamefont {Ablikim}} \emph {et~al.} (\bibinfo {collaboration} {BESIII Collaboration}),\ }\href {\doibase 10.1103/PhysRevD.102.031101} {\bibfield  {journal} {\bibinfo  {journal} {Phys. Rev. D}\ }\textbf {\bibinfo {volume} {102}},\ \bibinfo {pages} {031101} (\bibinfo {year} {2020}{\natexlab{b}})}\BibitemShut {NoStop}%
\bibitem [{\citenamefont {Ablikim}\ \emph {et~al.}(2024)\citenamefont {Ablikim} \emph {et~al.}}]{PRD109092012}%
  \BibitemOpen
  \bibfield  {author} {\bibinfo {author} {\bibfnamefont {M.}~\bibnamefont {Ablikim}} \emph {et~al.} (\bibinfo {collaboration} {BESIII Collaboration}),\ }\href {\doibase 10.1103/PhysRevD.109.092012} {\bibfield  {journal} {\bibinfo  {journal} {Phys. Rev. D}\ }\textbf {\bibinfo {volume} {109}},\ \bibinfo {pages} {092012} (\bibinfo {year} {2024})}\BibitemShut {NoStop}%
\bibitem [{\citenamefont {Ablikim}\ \emph {et~al.}(2022{\natexlab{b}})\citenamefont {Ablikim} \emph {et~al.}}]{PRL129102003}%
  \BibitemOpen
  \bibfield  {author} {\bibinfo {author} {\bibfnamefont {M.}~\bibnamefont {Ablikim}} \emph {et~al.} (\bibinfo {collaboration} {BESIII Collaboration}),\ }\href {\doibase 10.1103/PhysRevLett.129.102003} {\bibfield  {journal} {\bibinfo  {journal} {Phys. Rev. Lett.}\ }\textbf {\bibinfo {volume} {129}},\ \bibinfo {pages} {102003} (\bibinfo {year} {2022}{\natexlab{b}})}\BibitemShut {NoStop}%
\bibitem [{\citenamefont {Ablikim}\ \emph {et~al.}(2018)\citenamefont {Ablikim} \emph {et~al.}}]{PRD97052001}%
  \BibitemOpen
  \bibfield  {author} {\bibinfo {author} {\bibfnamefont {M.}~\bibnamefont {Ablikim}} \emph {et~al.} (\bibinfo {collaboration} {BESIII Collaboration}),\ }\href {\doibase 10.1103/PhysRevD.97.052001} {\bibfield  {journal} {\bibinfo  {journal} {Phys. Rev. D}\ }\textbf {\bibinfo {volume} {97}},\ \bibinfo {pages} {052001} (\bibinfo {year} {2018})}\BibitemShut {NoStop}%
\bibitem [{\citenamefont {Ablikim}\ \emph {et~al.}(2010{\natexlab{a}})\citenamefont {Ablikim} \emph {et~al.}}]{NIMA614345399}%
  \BibitemOpen
  \bibfield  {author} {\bibinfo {author} {\bibfnamefont {M.}~\bibnamefont {Ablikim}} \emph {et~al.} (\bibinfo {collaboration} {BESIII Collaboration}),\ }\href {\doibase 10.1016/j.nima.2009.12.050} {\bibfield  {journal} {\bibinfo  {journal} {Nucl. Instrum. Methods Phys. Res., Sect. A}\ }\textbf {\bibinfo {volume} {614}},\ \bibinfo {pages} {345} (\bibinfo {year} {2010}{\natexlab{a}})}\BibitemShut {NoStop}%
\bibitem [{\citenamefont {Yu}\ \emph {et~al.}(2016)\citenamefont {Yu} \emph {et~al.}}]{7IPACIPAC20162016}%
  \BibitemOpen
  \bibfield  {author} {\bibinfo {author} {\bibfnamefont {C.}~\bibnamefont {Yu}} \emph {et~al.},\ }in\ \href {\doibase 10.18429/JACoW-IPAC2016-TUYA01} {\emph {\bibinfo {booktitle} {{7th International Particle Accelerator Conference}}}},\ Vol.\ \bibinfo {volume} {IPAC2016}\ (\bibinfo  {publisher} {JACoW, Geneva, Switzerland},\ \bibinfo {year} {2016})\BibitemShut {NoStop}%
\bibitem [{\citenamefont {Ablikim}\ \emph {et~al.}(2020{\natexlab{c}})\citenamefont {Ablikim} \emph {et~al.}}]{CPC44040001}%
  \BibitemOpen
  \bibfield  {author} {\bibinfo {author} {\bibfnamefont {M.}~\bibnamefont {Ablikim}} \emph {et~al.} (\bibinfo {collaboration} {BESIII Collaboration}),\ }\href {\doibase 10.1088/1674-1137/44/4/040001} {\bibfield  {journal} {\bibinfo  {journal} {Chin. Phys. C}\ }\textbf {\bibinfo {volume} {44}},\ \bibinfo {pages} {040001} (\bibinfo {year} {2020}{\natexlab{c}})}\BibitemShut {NoStop}%
\bibitem [{\citenamefont {Lu}\ \emph {et~al.}(2020)\citenamefont {Lu}, \citenamefont {Xiao},\ and\ \citenamefont {Ji}}]{RDTaM4337344}%
  \BibitemOpen
  \bibfield  {author} {\bibinfo {author} {\bibfnamefont {J.}~\bibnamefont {Lu}}, \bibinfo {author} {\bibfnamefont {Y.}~\bibnamefont {Xiao}}, \ and\ \bibinfo {author} {\bibfnamefont {X.}~\bibnamefont {Ji}},\ }\href {\doibase 10.1007/s41605-020-00188-8} {\bibfield  {journal} {\bibinfo  {journal} {Radiat. Detect. Technol. Methods}\ }\textbf {\bibinfo {volume} {4}},\ \bibinfo {pages} {337} (\bibinfo {year} {2020})}\BibitemShut {NoStop}%
\bibitem [{\citenamefont {Zhang}\ \emph {et~al.}(2022)\citenamefont {Zhang} \emph {et~al.}}]{RDTaM6289293}%
  \BibitemOpen
  \bibfield  {author} {\bibinfo {author} {\bibfnamefont {J.~W.}\ \bibnamefont {Zhang}} \emph {et~al.},\ }\href {\doibase 10.1007/s41605-022-00331-7} {\bibfield  {journal} {\bibinfo  {journal} {Radiat. Detect. Technol. Methods}\ }\textbf {\bibinfo {volume} {6}},\ \bibinfo {pages} {289} (\bibinfo {year} {2022})}\BibitemShut {NoStop}%
\bibitem [{\citenamefont {Li}\ \emph {et~al.}(2017)\citenamefont {Li} \emph {et~al.}}]{RDTM1}%
  \BibitemOpen
  \bibfield  {author} {\bibinfo {author} {\bibfnamefont {X.}~\bibnamefont {Li}} \emph {et~al.},\ }\href {https://doi.org/10.1007/s41605-017-0014-2} {\bibfield  {journal} {\bibinfo  {journal} {Radiat. Detect. Technol. Methods}\ }\textbf {\bibinfo {volume} {1}} (\bibinfo {year} {2017})}\BibitemShut {NoStop}%
\bibitem [{\citenamefont {Guo}\ \emph {et~al.}(2017)\citenamefont {Guo} \emph {et~al.}}]{RDTM115}%
  \BibitemOpen
  \bibfield  {author} {\bibinfo {author} {\bibfnamefont {Y.~X.}\ \bibnamefont {Guo}} \emph {et~al.},\ }\href {\doibase 10.1007/s41605-017-0012-4} {\bibfield  {journal} {\bibinfo  {journal} {Radiat. Detect. Technol. Methods}\ }\textbf {\bibinfo {volume} {1}},\ \bibinfo {pages} {15} (\bibinfo {year} {2017})}\BibitemShut {NoStop}%
\bibitem [{\citenamefont {Cao}\ \emph {et~al.}(2020)\citenamefont {Cao} \emph {et~al.}}]{NIaMiPRSAASDaAE953163053}%
  \BibitemOpen
  \bibfield  {author} {\bibinfo {author} {\bibfnamefont {P.}~\bibnamefont {Cao}} \emph {et~al.},\ }\href {\doibase 10.1016/j.nima.2019.163053} {\bibfield  {journal} {\bibinfo  {journal} {Nucl. Instrum. Methods Phys. Res., Sect. A}\ }\textbf {\bibinfo {volume} {953}},\ \bibinfo {pages} {163053} (\bibinfo {year} {2020})}\BibitemShut {NoStop}%
\bibitem [{\citenamefont {Agostinelli}\ \emph {et~al.}(2003)\citenamefont {Agostinelli} \emph {et~al.}}]{NIMA506250303}%
  \BibitemOpen
  \bibfield  {author} {\bibinfo {author} {\bibfnamefont {S.}~\bibnamefont {Agostinelli}} \emph {et~al.} (\bibinfo {collaboration} {GEANT4 Collaboration}),\ }\href {\doibase 10.1016/S0168-9002(03)01368-8} {\bibfield  {journal} {\bibinfo  {journal} {Nucl. Instrum. Meth. A}\ }\textbf {\bibinfo {volume} {506}},\ \bibinfo {pages} {250} (\bibinfo {year} {2003})}\BibitemShut {NoStop}%
\bibitem [{\citenamefont {Jadach}\ \emph {et~al.}(2001)\citenamefont {Jadach}, \citenamefont {Ward},\ and\ \citenamefont {W{\c a}s}}]{PRD63113009}%
  \BibitemOpen
  \bibfield  {author} {\bibinfo {author} {\bibfnamefont {S.}~\bibnamefont {Jadach}}, \bibinfo {author} {\bibfnamefont {B.~F.~L.}\ \bibnamefont {Ward}}, \ and\ \bibinfo {author} {\bibfnamefont {Z.}~\bibnamefont {W{\c a}s}},\ }\href {\doibase 10.1103/PhysRevD.63.113009} {\bibfield  {journal} {\bibinfo  {journal} {Phys. Rev. D}\ }\textbf {\bibinfo {volume} {63}},\ \bibinfo {pages} {113009} (\bibinfo {year} {2001})}\BibitemShut {NoStop}%
\bibitem [{\citenamefont {Lange}(2001)}]{NIMA462152155}%
  \BibitemOpen
  \bibfield  {author} {\bibinfo {author} {\bibfnamefont {D.}~\bibnamefont {Lange}},\ }\href {\doibase 10.1016/S0168-9002(01)00089-4} {\bibfield  {journal} {\bibinfo  {journal} {Nucl. Instrum. Meth. A}\ }\textbf {\bibinfo {volume} {462}},\ \bibinfo {pages} {152} (\bibinfo {year} {2001})}\BibitemShut {NoStop}%
\bibitem [{\citenamefont {Ping}(2008)}]{CPC32599}%
  \BibitemOpen
  \bibfield  {author} {\bibinfo {author} {\bibfnamefont {R.~G.}\ \bibnamefont {Ping}},\ }\href {\doibase 10.1088/1674-1137/32/8/001} {\bibfield  {journal} {\bibinfo  {journal} {Chin. Phys. C}\ }\textbf {\bibinfo {volume} {32}},\ \bibinfo {pages} {599} (\bibinfo {year} {2008})}\BibitemShut {NoStop}%
\bibitem [{\citenamefont {Navas}\ \emph {et~al.}(2024)\citenamefont {Navas} \emph {et~al.}}]{PRD110030001}%
  \BibitemOpen
  \bibfield  {author} {\bibinfo {author} {\bibfnamefont {S.}~\bibnamefont {Navas}} \emph {et~al.} (\bibinfo {collaboration} {Particle Data Group}),\ }\href {\doibase 10.1103/PhysRevD.110.030001} {\bibfield  {journal} {\bibinfo  {journal} {Phys. Rev. D}\ }\textbf {\bibinfo {volume} {110}},\ \bibinfo {pages} {030001} (\bibinfo {year} {2024})}\BibitemShut {NoStop}%
\bibitem [{\citenamefont {Chen}\ \emph {et~al.}(2000)\citenamefont {Chen}, \citenamefont {Huang}, \citenamefont {Qi}, \citenamefont {Zhang},\ and\ \citenamefont {Zhu}}]{PRD62034003}%
  \BibitemOpen
  \bibfield  {author} {\bibinfo {author} {\bibfnamefont {J.~C.}\ \bibnamefont {Chen}}, \bibinfo {author} {\bibfnamefont {G.~S.}\ \bibnamefont {Huang}}, \bibinfo {author} {\bibfnamefont {X.~R.}\ \bibnamefont {Qi}}, \bibinfo {author} {\bibfnamefont {D.~H.}\ \bibnamefont {Zhang}}, \ and\ \bibinfo {author} {\bibfnamefont {Y.~S.}\ \bibnamefont {Zhu}},\ }\href {\doibase 10.1103/PhysRevD.62.034003} {\bibfield  {journal} {\bibinfo  {journal} {Phys. Rev. D}\ }\textbf {\bibinfo {volume} {62}},\ \bibinfo {pages} {034003} (\bibinfo {year} {2000})}\BibitemShut {NoStop}%
\bibitem [{\citenamefont {Yang}\ \emph {et~al.}(2014)\citenamefont {Yang}, \citenamefont {Ping},\ and\ \citenamefont {Chen}}]{CPL31061301}%
  \BibitemOpen
  \bibfield  {author} {\bibinfo {author} {\bibfnamefont {R.~L.}\ \bibnamefont {Yang}}, \bibinfo {author} {\bibfnamefont {R.~G.}\ \bibnamefont {Ping}}, \ and\ \bibinfo {author} {\bibfnamefont {H.}~\bibnamefont {Chen}},\ }\href {\doibase 10.1088/0256-307X/31/6/061301} {\bibfield  {journal} {\bibinfo  {journal} {Chin. Phys. C}\ }\textbf {\bibinfo {volume} {31}},\ \bibinfo {pages} {061301} (\bibinfo {year} {2014})}\BibitemShut {NoStop}%
\bibitem [{\citenamefont {Barberio}\ \emph {et~al.}(1991)\citenamefont {Barberio}, \citenamefont {van Eijk},\ and\ \citenamefont {W{\c a}s}}]{CPC66115128}%
  \BibitemOpen
  \bibfield  {author} {\bibinfo {author} {\bibfnamefont {E.}~\bibnamefont {Barberio}}, \bibinfo {author} {\bibfnamefont {B.}~\bibnamefont {van Eijk}}, \ and\ \bibinfo {author} {\bibfnamefont {Z.}~\bibnamefont {W{\c a}s}},\ }\href {\doibase 10.1016/0010-4655(91)90012-A} {\bibfield  {journal} {\bibinfo  {journal} {Comput. Phys. Commun.}\ }\textbf {\bibinfo {volume} {66}},\ \bibinfo {pages} {115} (\bibinfo {year} {1991})}\BibitemShut {NoStop}%
\bibitem [{\citenamefont {Actis}\ \emph {et~al.}(2010)\citenamefont {Actis} \emph {et~al.}}]{EPJC66585686}%
  \BibitemOpen
  \bibfield  {author} {\bibinfo {author} {\bibfnamefont {S.}~\bibnamefont {Actis}} \emph {et~al.},\ }\href {\doibase 10.1140/epjc/s10052-010-1251-4} {\bibfield  {journal} {\bibinfo  {journal} {Eur. Phys. J. C}\ }\textbf {\bibinfo {volume} {66}},\ \bibinfo {pages} {585} (\bibinfo {year} {2010})}\BibitemShut {NoStop}%
\bibitem [{\citenamefont {Sun}\ \emph {et~al.}(2021)\citenamefont {Sun}, \citenamefont {Liu}, \citenamefont {Jing}, \citenamefont {Wang}, \citenamefont {Zhong},\ and\ \citenamefont {Song}}]{FPB1664501}%
  \BibitemOpen
  \bibfield  {author} {\bibinfo {author} {\bibfnamefont {W.}~\bibnamefont {Sun}}, \bibinfo {author} {\bibfnamefont {T.}~\bibnamefont {Liu}}, \bibinfo {author} {\bibfnamefont {M.}~\bibnamefont {Jing}}, \bibinfo {author} {\bibfnamefont {L.}~\bibnamefont {Wang}}, \bibinfo {author} {\bibfnamefont {B.}~\bibnamefont {Zhong}}, \ and\ \bibinfo {author} {\bibfnamefont {W.}~\bibnamefont {Song}},\ }\href {\doibase 10.1007/s11467-021-1085-6} {\bibfield  {journal} {\bibinfo  {journal} {Front. Phys. (Beijing)}\ }\textbf {\bibinfo {volume} {16}},\ \bibinfo {pages} {64501} (\bibinfo {year} {2021})}\BibitemShut {NoStop}%
\bibitem [{\citenamefont {D'Agostini}(2003)}]{RoPiP661383}%
  \BibitemOpen
  \bibfield  {author} {\bibinfo {author} {\bibfnamefont {G.}~\bibnamefont {D'Agostini}},\ }\href {\doibase 10.1088/0034-4885/66/9/201} {\bibfield  {journal} {\bibinfo  {journal} {Reports on Progress in Physics}\ }\textbf {\bibinfo {volume} {66}},\ \bibinfo {pages} {1383} (\bibinfo {year} {2003})}\BibitemShut {NoStop}%
\bibitem [{\citenamefont {Ablikim}\ \emph {et~al.}(2022{\natexlab{c}})\citenamefont {Ablikim} \emph {et~al.}}]{CPC46113003}%
  \BibitemOpen
  \bibfield  {author} {\bibinfo {author} {\bibfnamefont {M.}~\bibnamefont {Ablikim}} \emph {et~al.} (\bibinfo {collaboration} {BESIII Collaboration}),\ }\href {\doibase 10.1088/1674-1137/ac84cc} {\bibfield  {journal} {\bibinfo  {journal} {Chin. Phys. C}\ }\textbf {\bibinfo {volume} {46}},\ \bibinfo {pages} {113003} (\bibinfo {year} {2022}{\natexlab{c}})}\BibitemShut {NoStop}%
\bibitem [{\citenamefont {Ablikim}\ \emph {et~al.}(2015{\natexlab{b}})\citenamefont {Ablikim} \emph {et~al.}}]{CPC39093001}%
  \BibitemOpen
  \bibfield  {author} {\bibinfo {author} {\bibfnamefont {M.}~\bibnamefont {Ablikim}} \emph {et~al.} (\bibinfo {collaboration} {BESIII Collaboration}),\ }\href {\doibase 10.1088/1674-1137/39/9/093001} {\bibfield  {journal} {\bibinfo  {journal} {Chin. Phys. C}\ }\textbf {\bibinfo {volume} {39}},\ \bibinfo {pages} {093001} (\bibinfo {year} {2015}{\natexlab{b}})}\BibitemShut {NoStop}%
\bibitem [{\citenamefont {Ablikim}\ \emph {et~al.}(2022{\natexlab{d}})\citenamefont {Ablikim} \emph {et~al.}}]{CPC46113002}%
  \BibitemOpen
  \bibfield  {author} {\bibinfo {author} {\bibfnamefont {M.}~\bibnamefont {Ablikim}} \emph {et~al.} (\bibinfo {collaboration} {BESIII Collaboration}),\ }\href {\doibase 10.1088/1674-1137/ac80b4} {\bibfield  {journal} {\bibinfo  {journal} {Chin. Phys. C}\ }\textbf {\bibinfo {volume} {46}},\ \bibinfo {pages} {113002} (\bibinfo {year} {2022}{\natexlab{d}})}\BibitemShut {NoStop}%
\bibitem [{\citenamefont {Ablikim}\ \emph {et~al.}(2013)\citenamefont {Ablikim} \emph {et~al.}}]{PRL110252001}%
  \BibitemOpen
  \bibfield  {author} {\bibinfo {author} {\bibfnamefont {M.}~\bibnamefont {Ablikim}} \emph {et~al.} (\bibinfo {collaboration} {BESIII Collaboration}),\ }\href {\doibase 10.1103/PhysRevLett.110.252001} {\bibfield  {journal} {\bibinfo  {journal} {Phys. Rev. Lett.}\ }\textbf {\bibinfo {volume} {110}},\ \bibinfo {pages} {252001} (\bibinfo {year} {2013})}\BibitemShut {NoStop}%
\bibitem [{\citenamefont {Ablikim}\ \emph {et~al.}(2011)\citenamefont {Ablikim} \emph {et~al.}}]{PRD83112005}%
  \BibitemOpen
  \bibfield  {author} {\bibinfo {author} {\bibfnamefont {M.}~\bibnamefont {Ablikim}} \emph {et~al.} (\bibinfo {collaboration} {BESIII Collaboration}),\ }\href {\doibase 10.1103/PhysRevD.83.112005} {\bibfield  {journal} {\bibinfo  {journal} {Phys. Rev. D}\ }\textbf {\bibinfo {volume} {83}},\ \bibinfo {pages} {112005} (\bibinfo {year} {2011})}\BibitemShut {NoStop}%
\bibitem [{\citenamefont {Ablikim}\ \emph {et~al.}(2010{\natexlab{b}})\citenamefont {Ablikim} \emph {et~al.}}]{PRD81052005}%
  \BibitemOpen
  \bibfield  {author} {\bibinfo {author} {\bibfnamefont {M.}~\bibnamefont {Ablikim}} \emph {et~al.} (\bibinfo {collaboration} {BESIII Collaboration}),\ }\href {\doibase 10.1103/PhysRevD.81.052005} {\bibfield  {journal} {\bibinfo  {journal} {Phys. Rev. D}\ }\textbf {\bibinfo {volume} {81}},\ \bibinfo {pages} {052005} (\bibinfo {year} {2010}{\natexlab{b}})}\BibitemShut {NoStop}%
\bibitem [{\citenamefont {Ping}(2014)}]{CPC38083001}%
  \BibitemOpen
  \bibfield  {author} {\bibinfo {author} {\bibfnamefont {R.~G.}\ \bibnamefont {Ping}},\ }\href {\doibase 10.1088/1674-1137/38/8/083001} {\bibfield  {journal} {\bibinfo  {journal} {Chin. Phys. C}\ }\textbf {\bibinfo {volume} {38}},\ \bibinfo {pages} {083001} (\bibinfo {year} {2014})}\BibitemShut {NoStop}%
\bibitem [{\citenamefont {Barlow}(2002)}]{arXivhepex0207026}%
  \BibitemOpen
  \bibfield  {author} {\bibinfo {author} {\bibfnamefont {R.}~\bibnamefont {Barlow}},\ }\href {\doibase 10.48550/arXiv.hep-ex/0207026} {\enquote {\bibinfo {title} {Systematic {Errors}: facts and fictions},}\ } (\bibinfo {year} {2002}),\ \bibinfo {note} {arXiv:hep-ex/0207026}\BibitemShut {NoStop}%
\bibitem [{\citenamefont {Ablikim}\ \emph {et~al.}(2022{\natexlab{e}})\citenamefont {Ablikim} \emph {et~al.}}]{N6066469}%
  \BibitemOpen
  \bibfield  {author} {\bibinfo {author} {\bibfnamefont {M.}~\bibnamefont {Ablikim}} \emph {et~al.} (\bibinfo {collaboration} {BESIII Collaboration}),\ }\href {\doibase 10.1038/s41586-022-04624-1} {\bibfield  {journal} {\bibinfo  {journal} {Nature}\ }\textbf {\bibinfo {volume} {606}},\ \bibinfo {pages} {64} (\bibinfo {year} {2022}{\natexlab{e}})}\BibitemShut {NoStop}%
\bibitem [{\citenamefont {Ablikim}\ \emph {et~al.}(2016)\citenamefont {Ablikim} \emph {et~al.}}]{CPC40063001}%
  \BibitemOpen
  \bibfield  {author} {\bibinfo {author} {\bibfnamefont {M.}~\bibnamefont {Ablikim}} \emph {et~al.} (\bibinfo {collaboration} {BESIII Collaboration}),\ }\href {\doibase 10.1088/1674-1137/40/6/063001} {\bibfield  {journal} {\bibinfo  {journal} {Chin. Phys. C}\ }\textbf {\bibinfo {volume} {40}},\ \bibinfo {pages} {063001} (\bibinfo {year} {2016})}\BibitemShut {NoStop}%
\bibitem [{\citenamefont {Ablikim}\ \emph {et~al.}(2021{\natexlab{b}})\citenamefont {Ablikim} \emph {et~al.}}]{CPC45103001}%
  \BibitemOpen
  \bibfield  {author} {\bibinfo {author} {\bibfnamefont {M.}~\bibnamefont {Ablikim}} \emph {et~al.} (\bibinfo {collaboration} {BESIII Collaboration}),\ }\href {\doibase 10.1088/1674-1137/ac1575} {\bibfield  {journal} {\bibinfo  {journal} {Chin. Phys. C}\ }\textbf {\bibinfo {volume} {45}},\ \bibinfo {pages} {103001} (\bibinfo {year} {2021}{\natexlab{b}})}\BibitemShut {NoStop}%
\bibitem [{\citenamefont {Abakumova}\ \emph {et~al.}(2011)\citenamefont {Abakumova} \emph {et~al.}}]{NIMA6592129}%
  \BibitemOpen
  \bibfield  {author} {\bibinfo {author} {\bibfnamefont {E.~V.}\ \bibnamefont {Abakumova}} \emph {et~al.},\ }\href {\doibase 10.1016/j.nima.2011.08.050} {\bibfield  {journal} {\bibinfo  {journal} {Nucl. Instrum. Methods Phys. Res., Sect. A}\ }\textbf {\bibinfo {volume} {659}},\ \bibinfo {pages} {21} (\bibinfo {year} {2011})}\BibitemShut {NoStop}%
\end{thebibliography}
%

\end{document}